\documentclass[a4paper,11pt]{article}
\usepackage{pos}

\usepackage[utf8]{inputenc} 

\usepackage{pmboxdraw} 

\usepackage[T1]{fontenc}    
\usepackage{hyperref}       
\usepackage{url}            
\usepackage{booktabs}       
\usepackage{amsfonts}       
\usepackage{nicefrac}       
\usepackage{microtype}      
\usepackage{lipsum}

\usepackage{amsmath}
\usepackage{bm}
\usepackage{bbm}
\usepackage{amsfonts}
\usepackage{graphicx}

\usepackage{colortbl}
\usepackage{tikz}

\graphicspath{{./figs}}

\usetikzlibrary{decorations.pathmorphing}
\usetikzlibrary{shapes}
\usetikzlibrary{shapes.geometric}
\usetikzlibrary{decorations.pathreplacing}
\usepackage[force]{feynmp-auto}
\usepackage{tabularx}
\usepackage{pbox}
\usepackage{slashed}
\usepackage{braket}
\usepackage{hyperref}
\hypersetup{
    colorlinks=true,
    linkcolor=blue,
    filecolor=magenta,      
    urlcolor=blue,
}

\usepackage{listings}
\usepackage{xcolor}

\definecolor{codegreen}{rgb}{0,0.6,0}
\definecolor{codegray}{rgb}{0.5,0.5,0.5}
\definecolor{codepurple}{rgb}{0.58,0,0.82}
\definecolor{backcolour}{rgb}{0.95,0.95,0.92}

\lstdefinestyle{mystyle}{
    backgroundcolor=\color{backcolour},   
    commentstyle=\color{codegreen},
    keywordstyle=\color{magenta},
    numberstyle=\tiny\color{codegray},
    stringstyle=\color{codepurple},
    basicstyle=\ttfamily\footnotesize,
    breakatwhitespace=false,         
    breaklines=true,                 
    captionpos=b,                    
    keepspaces=true,                 
    numbers=left,                    
    numbersep=5pt,                  
    showspaces=false,                
    showstringspaces=false,
    showtabs=false,                  
    tabsize=2
}

\newcommand{\sigx}{\begin{pmatrix} 0 & 1 \\ 1 & 0 \end{pmatrix}}
\newcommand{\sigy}{\begin{pmatrix} 0 & -\bm{i} \\ \bm{i} & 0 \end{pmatrix}}
\newcommand{\sigz}{\begin{pmatrix} 1 & 0 \\ 0 & -1 \end{pmatrix}}
\newcommand{\sigp}{\begin{pmatrix} 0 & 1 \\ 0 & 0 \end{pmatrix}}
\newcommand{\sigm}{\begin{pmatrix} 0 & 0 \\ 1 & 0 \end{pmatrix}}
\newcommand{\proja}{\begin{pmatrix} 1 & 0 \\ 0 & 0 \end{pmatrix}}
\newcommand{\projb}{\begin{pmatrix} 0 & 0 \\ 0 & 1 \end{pmatrix}}

\title{Quantum Information for Particle Theorists}

\author{Joseph D.~Lykken}

\affiliation{Fermilab Quantum Institute
and Theoretical Physics Department,\\ 
Fermi National Accelerator Laboratory\\
Batavia, IL 60510, USA}

\emailAdd{lykken@fnal.gov}

\abstract{
Lectures given at the Theoretical Advanced Study Institute (TASI 2020), 1-26 June 2020. The topics covered include quantum circuits, entanglement,
quantum teleportation, Bell inequalities, quantum entropy and decoherence, classical versus quantum measurement, the area law for entanglement
entropy in quantum field theory, and simulating quantum field theory on a quantum computer. 
Along the way we confront the fundamental
sloppiness of how we all learned (and some of us taught) quantum mechanics in college.
Links to a Python notebook and Mathematica notebooks \cite{notebooks}
will allow the reader to reproduce and extend the calculations, as well as perform five experiments on a quantum simulator.
\\
\\
FERMILAB-CONF-20-502-QIS-T
}

\FullConference{%
  TASI 2020 -- The Obscure Universe: Neutrinos and Other Dark Matters\\
  1-26 June, 2020\\
  Boulder, Colorado, USA}

\tableofcontents

\begin{document}
\maketitle

\section{Introduction: the overlap between particle physics and quantum information science}\label{sec:one}

Particle physics has the ambitious goals of uncovering the most fundamental constituents of reality and deciphering the rules by which
those constituents interact. Those rules include quantum mechanics, and the fundamental constituents appear to be quantum entities.
For example in the Standard Model we talk about excitations of relativistic quantum fields that are characterized by fixed quantum numbers such as
mass, spin, and various charges. Furthermore in particle physics experiments we have the capability to produce states that are quantum
superpositions of certain quantum numbers. For example (muon) neutrinos produced from pion decays in various beams at Fermilab are in
a quantum superposition of (at least) three different neutrino mass eigenstates, and that superposition changes over time as a result of the
usual quantum unitary time evolution represented by the operator exp$(-iHt)$, where $H$ is the neutrino Hamiltonian.
Thus neutrino oscillation experiments are an example of studying the time evolution of {\it quantum information} on macroscopic scales.

The NOvA long baseline neutrino experiment is sensitive to another effect well studied in quantum information science (QIS): as the neutrinos
travel 800 kilometers through the Earth from Fermilab to Ash River, Minnesota, they are performing (via charged current weak interactions) a
{\it quantum nondemolition measurement} of the number density of electrons in the Earth's crust; this measurement is in the form of 
a frequency shift that affects the fraction of the neutrino superposition that corresponds to an electron neutrino flavor
eigenstate. 

{\it Quantum entanglement} is fundamental to both QIS and particle physics, but in the latter case is usually discussed in the language of quantum correlations.
Consider the discovery of the Higgs boson at the LHC; one of the two discovery channels was looking for the process where a Higgs boson is produced in a
proton-proton collision and then decays weakly via two virtual $Z$ bosons to four charged leptons (electrons or muons). In the approximation that the leptons
are massless, the measured final state has 16 observables (the three-momenta of the four leptons and their electric charges). These observables have quantum correlations that arise
from the fact that the four leptons come from a Higgs boson decay and are thus entangled, meaning that the Higgs parent state cannot be decomposed as a
tensor product of single lepton decay states. Thus for example the 16 observables are correlated such that the total invariant mass computed from them is
strongly peaked at 125 GeV$/c^2$; there are also angular correlations that trace back to the fact that Higgs is a spinless even parity boson.
These correlations, arising from quantum entanglement, were used in the CMS discovery analysis to help distinguish the Higgs signal from backgrounds;
had we not so used our understanding of entanglement, there would not have been a CMS discovery announcement on July 4, 2012.

Understanding the initial state strong dynamics of proton-proton collisions at the LHC may eventually be an outstanding example of following
Richard Feynman's dictum to use quantum computers to solve quantum problems. The underlying theory is known: it is QCD. In that theory we
know how to formulate the problem in terms of the partons of one proton probing the other proton; this is a messier version of deep inelastic
scattering, where we describe a colorless particle probing a proton. The process involves details of real-time strong dynamics, and is thus not amenable to
the approach of Euclidean-time lattice gauge theory\footnote{As an example of how one would formulate this in QCD, look at equation 4.42 and the
accompanying text in Ellis, Stirling, and Webber's book \cite{Ellis:1991qj}.}
The current state of the art is to use data-based parameterizations to describe both parton
distributions and the processes of fragmentation and hadronization. This is somewhat embarrassing more than forty years after developing QCD.
Quantum computers, by construction, efficiently compute the real-time dynamics of strongly interacting quantum systems. While it may take a long time
for quantum computers to be sufficiently performant to make important contributions to hadronic physics, a lot of progress has already been made in
thinking about how to map particle physics dynamics into quantum algorithms.

One important topic that I will not cover here is the fascinating relationship between quantum information, black holes, and holography as seen via AdS/CFT
duality. This is a huge area of active research that has also a long history; indeed the first paper that I ever read that discusses entanglement entropy is
Don Page's classic 1993 paper ``Information in black hole radiation"  \cite{Page:1993wv}. It would not surprise me if this line of research eventually leads
to concrete experimentally testable proposals for emergent spacetime and emergent gravity in laboratory quantum systems. Indeed Gao and Jafferis
have already published a concrete recipe for a quantum simulation that should exhibit some of the properties of a traversable wormhole \cite{Gao:2019nyj}. 
An excellent introduction to all
this are the TASI and Jerusalem lectures by Dan Harlow \cite{Harlow:2018fse,Harlow:2014yka}.

\subsection{Outline}
In Section \ref{sec:two}, I introduce the basic theoretical concepts of quantum computing, in the language of {\it qubits} and {\it entanglement}.
We will perform three simple experiments with {\it quantum circuits} constructed using {\it quantum gates} that act as
unitary operators on one or two qubits at a time. In the Python notebook provided you can run all of these circuits on the Google Cirq
quantum simulator. In Section \ref{sec:three} we extend this to perform two experiments with {\it quantum teleportation}.

In Section \ref{sec:four}, I introduce the basic concepts of classical information theory and remind you of the basic techniques of classical measurement
as used, for example, in experiments at the LHC.
This language will give us enough knowledge to then discuss in Section \ref{sec:five} the general phenomena
of {\it quantum decoherence}, which in turn leads in the next section to a modern discussion of {\it quantum measurement}.
Along the way we will confront the fundamental
sloppiness of how we all learned (and some of us taught) quantum mechanics in college. Quantum computers are not thought experiments; they actually exist
and they manipulate known quantum states to get known results. This encourages us as theorists to suppress our philosphical tendencies and think more concretely.
Personally I have found this to be an extremely valuable exercise; it has changed fundamentally how I think about quantum physics.

Section \ref{sec:Bellinequality} is a short introduction to the famous Bell inequalities.
Section \ref{sec:six} is devoted to some remarkable properties of quantum entanglement derived in quantum field theory, in particular
the area law of entanglement entropy. Finally, in Section \ref{sec:seven},
I will discuss some of the issues involved in trying to simulate quantum field theories on actual quantum computers.
This is a new and fast-moving area of active research, with lots of good opportunities for junior researchers to jump in and do something
impactful.

\section{Quantum circuits and entanglement}\label{sec:two}

A quantum state space is constructed from a complex vector space called
a Hilbert space, and in QIS we will always
assume it to have finite dimensionality $d$.
Quantum states are described by state vectors, which are just normed vectors in some Hilbert space modulo the fact that there is an irrelevant
overall phase. For $d=1$ we call this quantum state
a {\it qubit}; a qubit is just a normed vector in the complex projective space $\mathbb{CP}^1$.
Without loss of generality we can write a general qubit state as
\begin{eqnarray}
\ket{\psi} = {\rm cos}\frac{\theta}{2}\ket{0} + {\rm sin}\frac{\theta}{2}\,{\rm e}^{\phi} \ket{1}
\label{eq:qubit}
\end{eqnarray}
where $0 \leq \theta \leq \pi$, $0 \leq \phi \leq 2\pi$, and the kets $\ket{0}$, $\ket{1}$ are any convenient orthonormal basis.
Whatever this basis is in a particular physical realization, we will refer to it as the {\it computational basis}.
It is often useful to map this state to the surface of a sphere, called the {\it Bloch sphere}, treating $\theta$ as the polar angle
and $\phi$ as the azimuthal angle. In this language the basis states $\ket{0}$ , $\ket{1}$ are the North and South poles,
respectively. The antipodal points where the x-axis intersects the surface of the sphere correspond to the orthonormal states
\begin{eqnarray}
\ket{+} &\equiv& \frac{1}{\sqrt{2}}\left( \ket{0} + \ket{1} \right) \\
\ket{-} &\equiv& \frac{1}{\sqrt{2}}\left( \ket{0} - \ket{1} \right)
\label{eq:hadbasis}
\end{eqnarray}
These states are known as the {\it Hadamard basis}, since they can be obtained from the computational basis
by applying a $2\times 2$ unitary matrix called the Hadamard transformation $h$:
\begin{eqnarray}
\begin{pmatrix}
\ket{+} \\
\ket{-}
\end{pmatrix}
= 
h \cdot 
\begin{pmatrix}
\ket{0} \\
\ket{1}
\end{pmatrix}
=
\frac{1}{\sqrt{2}}
\begin{pmatrix}
1 & 1 \\
1 & -1 
\end{pmatrix}
\begin{pmatrix}
\ket{0} \\
\ket{1}
\end{pmatrix}
\label{eq:hadamard}
\end{eqnarray}
This is related to the properties of the Pauli matrices acting on a qubit state. Writing the Pauli matrices in the
computational basis
\begin{eqnarray}
\sigma^x = \sigx
\;,\quad
\sigma^y = \sigy
\;,\quad
\sigma^z = \sigz
\label{eq:Paulimatrices}
\end{eqnarray}
we see that $\sigma^z$ is diagonal with eigenvalues $\pm 1$ in the computational basis, while $\sigma^x$ is diagonal with eigenvalues
$\pm 1$ in the Hadamard basis.
Similarly the antipodal points where the y-axis intersects the surface of the sphere correspond to the orthnormal states
\begin{eqnarray}
\ket{\bm{i}} &\equiv& \frac{1}{\sqrt{2}}\left( \ket{0} +\bm{i} \ket{1} \right) \\
\ket{\bm{-i}} &\equiv& \frac{1}{\sqrt{2}}\left( \ket{0} -\bm{i} \ket{1} \right)
\label{eq:hadbasis}
\end{eqnarray}

%
\begin{figure}[tb]
\centering
\includegraphics[width=0.7\textwidth]{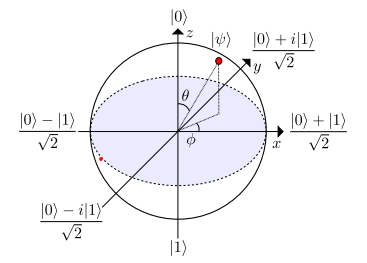}
\caption{The Bloch sphere representation of single qubit states. Taken from \cite{Kockum}.\label{fig:Bloch}}
\end{figure}
%

Already one might think that a state space constructed from qubits can carry a lot more information than an analogous state
space represented by classical binary bits, since the qubit state is specified in terms of two continuous real parameters $\theta$ and $\phi$.
To put it another way, the classical description only tells us whether we are in the northern hemisphere or the southern
hemisphere, while the qubit state vector distinguishes every point on the surface of the Bloch sphere. However we still have to confront
the notorious problem of quantum measurement. For now I will employ the canonical sloppy definition of quantum measurement
as a projection onto the computational basis states $\ket{0}$ and $\ket{1}$. Thus I will assume any particular measurement of an arbitary qubit
state always gives either 0 or 1 (this is equivalent to saying that we project to $\pm 1$, the eigenvalues of the Hermitian operator $\sigma^z$).
So measuring one qubit returns one classical bit.

You might say that there is still lots of wild and wonderful quantumness in the fact that the qubit state is in general a superposition of
the basis states $\ket{0}$ and $\ket{1}$. Thus if I prepare a large number $N$ of identical single qubit states, then measure them all,
I will get 0 approximately $N\,$cos$^2(\theta/2)$ times and get 1 approximately $N\,$sin$^2(\theta/2)$ times. The way probabilities
enter here is different from how they enter a classical description. Imagine a physical process (you can actually do this today on
a quantum computer) where you start in the state $\ket{0}$ and the first step of the process evolves the state by the unitary Hadamard
operator $h$. At this point classical reasoning would say that we have a 50\% chance of still being in the state $\ket{0}$ and 50\%
chance of flipping to $\ket{1}$. Now suppose that in the second step we evolve again by the Hadamard operator. Then the classical reasoning says
that if we were still in state $\ket{0}$ after the first step, there is a 50\% chance that we will stay in $\ket{0}$ after the second step;
furthermore if we happened to flip after the first step, there is a 50\% chance that we will flip again on the second step, ending up back in
$\ket{0}$. So classical reasoning says that after applying $h^2$ the total probability of being in state $\ket{0}$ is $(0.5)(0.5)+(0.5)(0.5) = 0.5$,
i.e. 50\%. However $h^2$ is just the identity matrix, so in fact this physical process returns the original state $\ket{0}$ 100\% of the time.
Of course the difference here is that the computation of probabilities arising from quantum measurement can involve interference effects. 
We can mock this up classically (with classical wave behavior), but quantum physics has it built in whether we like it or not.

One important lesson of the failure of the classical reasoning outlined above is that we should be careful in how we think about a
quantum superposition state. The qubit state $\ket{+} = (1/\sqrt{2})(\ket{0} + \ket{1})$ {\it is not} in both the $\ket{0}$ and
$\ket{1}$ state at the same time. Indeed it is not in {\it either one} of those states - it is instead in a different physical state, called the $\ket{+}$
state. Indeed in some actual physical realizations of qubits the Hadamard basis is just as good a basis for measuring states as the 
computational basis. If I switch the basis used to measure, then the $\ket{+}$ state is not a superposition state at all. Indeed for any
state in the Hilbert space, there is always a choice of measurement basis where that state is one of the basis states, and thus
not a superposition. So superposition does not really mean anything unless you find that some measurement bases are preferred over others;
we will return to this point when we discuss measurement again in Section \ref{sec:measure}.

Digital quantum computers execute programs as quantum circuits based on unitary gate operations on qubits.
A quantum computing circuit is a network (an acyclic graph) of {\it quantum logic gates}. Each gate 
performs some unitary transformation on one or more qubits. 

As with classical computing, quantum circuits are constructed using a small menu of different kinds of operations.
Here are some important examples. We begin with single qubit gates; obviously the action of any single qubit gate can be
represented by a rotation on the surface of the Bloch sphere around a particular axis by a particular amount.
\begin{itemize}

\item $\bm{X}$, $\bm{Y}$, $\bm{Z}$ gates:
An  $\bm{X}$ gate acts on a single qubit by interchanging the $\ket{0}$ and $\ket{1}$ basis states, thus the quantum equivalent of the Boolean NOT operation. This is a unitary transformation that in the computational basis
is just the Pauli matrix $\sigma^x$:
\begin{eqnarray} \label{eq:Xgate}
\bm{X} \equiv
\begin{pmatrix}
0 & 1\\ 
1 & 0
\end{pmatrix}
\end{eqnarray}
This is equivalent to a rotation by $\pi$ around the $x$-axis on the Bloch sphere.
We can also implement single qubit transformations $\bm{Y}$ and $\bm{Z}$
equivalent to the other two Pauli matrices  $\sigma^y$  and  $\sigma^z$. 
These can be exponentiated to generate arbitrary rotations around the $x$, $y$, and $z$ axes of the Bloch sphere.
For example:
\begin{eqnarray}
{\rm exp}\left(i\, \bm{X}\pi t/2\right) = {\rm cos}(\pi t/2) \mathbbm{1} + i\,{\rm sin}(\pi t/2) \bm{X} = \left(i\bm{X} \right)^t
\end{eqnarray}
The first equality in the above expression is easily checked by taking the powers series expansions of both sides. The second equality
is derived by noting that the first equality implies $i\bm{X} = {\rm exp}(i\, \bm{X}\pi /2)$.

Another common notation is the $\bm{R_{\varphi}}$ gate, which is equivalent to $\bm{Z}^{\varphi/\pi}$. This is an operator that
introduces a phase difference between the two computational basis states:
\begin{eqnarray}
\bm{R_{\varphi}} = 
\begin{pmatrix} 1 & 0 \\ 0 & {\rm e}^{\varphi} \end{pmatrix}
\end{eqnarray}
\item{\bf Hadamard gate}
The Hadamard gate $\bm{h}$ acts on the single qubit states like a rotation of $\pi$ around the $x + z$ axis.
 This is the following unitary transformation in the computational basis:
\begin{eqnarray} \label{eq:hgate}
\bm{h} = \frac{1}{\sqrt{2}}
\begin{pmatrix} 1 & 1 \\ 1 & -1 \end{pmatrix}
\end{eqnarray}
The Hadamard is usually labelled by a capital $H$, but I will reserve that to denote Hamiltonians.

\end{itemize}

It is conventional to start our quantum circuits
with all qubits in the state $\ket{0}$; we can then use single qubit gates to prepare more complicated superpositions. 
To do actual quantum computing we need to entangle the states of different qubits. It is not necessary
to have gates that fully entangle all the qubits in one operation; in the same way that the Boolean operations constituting a classical computing algorithm can be
composed of clauses each of which utilizes only a few bit registers,
quantum computers can do an arbitrary calculation using only gates that entangle two qubits at a time. In fact there are theorems to the effect all you need
to implement is one non-trivial two-qubit entangling gate, which is usually assumed to be the {\bf CNOT} gate. 
This is called a controlled NOT gate, since the state of one ``control" qubit controls whether the NOT operation is 
performed on the other ``target" qubit. We can write this as a unitary transformation in the $2^2 = 4$ state Hilbert space of two qubits; the computational basis of a pair of qubits
can be denoted by kets as $\ket{00}$, $\ket{01}$, $\ket{10}$, $\ket{11}$, or equivalently by the following vectors in the Hilbert space:
$(1000)$, $(0100)$, $(0010)$, $(0001)$. In the vector notation {\bf CNOT} acts as a $4\times 4$ unitary matrix; if the first qubit is the control
then:
\begin{eqnarray} \label{eq:CNOT1} 
{\bf CNOT} = 
\begin{pmatrix}
1 & 0 & 0 & 0 \\
0 & 1 & 0 & 0 \\
0 & 0 & 0 & 1 \\
0 & 0 & 1 & 0
\end{pmatrix}
\quad .
\end{eqnarray}
You should figure out on your own what the matrix is in the case where the second qubit is the control. Notice that the {\bf CNOT}
could just as well be called a controlled $\bm{X}$ operation or ${\bf CX}$. It is easy to define other kinds of controlled
gate operations; for example ${\bf CY}$, ${\bf CZ}$  perform a $\bm{Y}$ or $\bm{Z}$ operation on the target qubit
controlled by the state of the control qubit.


%
\begin{figure}[tb]
\centering
\includegraphics[width=0.6\textwidth]{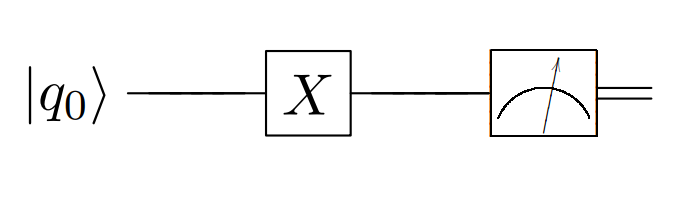}
\caption{Simple quantum circuit. The initial state is $\ket{0}$, the $\bm{X}$ gate flips this to the state $\ket{1}$,
and the dial indicates measurement in the computational basis.\label{fig:Exp1}}
\end{figure}
%


\subsection{Experiment 1: Simple quantum circuit} 

Let's run a very simple quantum circuit using the Cirq quantum simulator; the simulator is not a quantum computer, but it might as well be, since exactly
the same software does run on existing Google quantum computers. The circuit is shown in Figure~\ref{fig:Exp1}. It consists of a single qubit $q_0$
prepared initially in the $\ket{0}$ state. The circuit consists of acting with the unitary gate operation $\bm{X}$, then measuring the qubit again
in the computational basis.
With the notebook provided you can run this yourself using the following Python code:

\begin{lstlisting}[language=Python]
# Get a qubit and a circuit
qbit = cirq.LineQubit(0)
circuit = cirq.Circuit()

# Add an X gate: acts like the Pauli Matrix sigma_x
circuit.append(cirq.X(qbit))

# Run a simple simulation that extracts the wavefunction of this state
sim = cirq.Simulator()
result = sim.simulate(circuit)
printmd("\n**Bloch Sphere of the qubit in the final state:**")
state = cirq.bloch_vector_from_state_vector(result.final_state,0)
print("x: ", around(state[0], 4), " y: ", around(state[1], 4),
      " z: ", around(state[2], 4))

# Add a measurement at the end of the circuit: 
circuit.append(cirq.measure(qbit, key="Final state"))

# Display the circuit:
printmd("\n**Cirq circuit:**")
print(circuit)

# Invoke the Cirq quantum simulator to execute the circuit:
simulator = cirq.Simulator()

# Simulate the circuit several times:
result = simulator.run(circuit, repetitions=10)

# Print the results:
printmd("\n**Results of 10 trials:**")
print(result)
\end{lstlisting}
The output looks like this:
\begin{verbatim}
Bloch Sphere of the qubit in the final state:

x:  0.0  y:  0.0  z:  -1.0

Cirq circuit:

0: ───X───M('Final state')───

Results of 10 trials:

Final state=1111111111
\end{verbatim}

In this example we have used the nice feature of Cirq that in the simulator you can extract the actual quantum final state wavefunction, which of course you
cannot do in the real quantum computer, where you instead have to run the same circuit many times to figure out what the final state was.
In this simple example the final state is just $\ket{1}$.


\subsection{Entanglement and Bell states \label{ss:bell}}

For a system of two qubits, we can convert from the computational basis to an equivalent orthonormal basis defined by the four {\it Bell states}:

\begin{eqnarray} \label{sb}
|\beta_{00}\rangle &=& \frac{1}{\sqrt{2}}  \left(  |00\rangle + |11\rangle  \right) \nonumber\\
|\beta_{01}\rangle &=& \frac{1}{\sqrt{2}}  \left(  |01\rangle + |10\rangle \right) \nonumber\\
|\beta_{10}\rangle&=& \frac{1}{\sqrt{2}}  \left(  |00\rangle - |11\rangle \right) \nonumber\\
|\beta_{11}\rangle &=& \frac{1}{\sqrt{2}}  \left(  |01\rangle - |10\rangle  \right) 
\quad .
\end{eqnarray}

\begin{figure}[tb]
\centering
\includegraphics[width=0.7\textwidth]{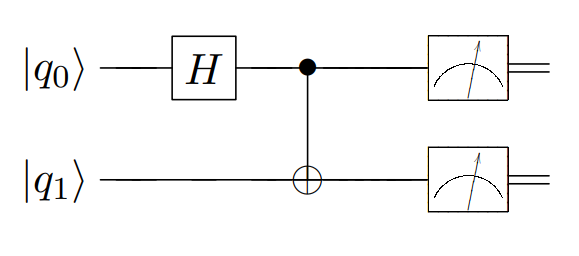}
\caption{Gate operations transform the initial state $\ket{00}$ into the state $\ket{\beta_{00}}$, which
represents an entangled EPR pair.\label{fig:Exp2}}
\end{figure}

Once you know how to make two-qubit states in the computational basis, you can convert them into Bell states by using a Hadamard gate on the first qubit followed by a CNOT gate on the pair of qubits, thus:
\begin{eqnarray} \label{sb}
|00\rangle \to |\beta_{00}\rangle &=& \frac{1}{\sqrt{2}}  \left(  |00\rangle + |11\rangle  \right) \nonumber\\
|01\rangle \to |\beta_{01}\rangle &=& \frac{1}{\sqrt{2}}  \left(  |01\rangle + |10\rangle \right)  \nonumber\\
|10\rangle \to |\beta_{10}\rangle &=& \frac{1}{\sqrt{2}}  \left(  |00\rangle - |11\rangle \right) \nonumber\\
|11\rangle \to |\beta_{11}\rangle &=& \frac{1}{\sqrt{2}}  \left(  |01\rangle - |10\rangle  \right) 
\quad .
\end{eqnarray}
A quantum circuit for this is illustrated in Figure~\ref{fig:Exp2}. This unitary transformation in the Hilbert space of two qubits is just a change of basis
from the two-qubit computational basis to the {\it Bell state basis}. If you perform the unitary operations in the reverse order, you can of course rotate
from the Bell basis back to the computational basis. This inverse operation, following by measurements of the qubits, is known as a {\it Bell state measurement},
since the net effect is to project your two-qubit state, whatever it was, onto one of the four Bell states.


%
%


\subsection{Experiment 2: Create a Bell state} 

Let's run the quantum circuit shown in Figure~\ref{fig:Exp2}. With the notebook provided you can run this on the Cirq quantum simulator using the following Python code:

\begin{lstlisting}[language=Python]
# Get two qubits and a circuit
qubit = [cirq.LineQubit(x) for x in range(2)]
circuit = cirq.Circuit()

# Add a Hadamard gate to qubit 0, then a CNOT gate from qubit 0 to qubit 1:
circuit.append([cirq.H(qubit[0]), 
             cirq.CNOT(qubit[0], qubit[1])])

# Run a simple simulation that extracts the actual final states
sim = cirq.Simulator()
result = sim.simulate(circuit)
printmd("\n**Bloch Sphere of the qubit 0 in the final state:**")
state = cirq.bloch_vector_from_state_vector(result.final_state,0)
print("x: ", around(state[0], 4), " y: ", around(state[1], 4),
      " z: ", around(state[2], 4))
printmd("\n**Bloch Sphere of the qubit 1 in the final state:**")
state = cirq.bloch_vector_from_state_vector(result.final_state,1)
print("x: ", around(state[0], 4), " y: ", around(state[1], 4),
      " z: ", around(state[2], 4))

# Add a measurement at the end of the circuit: 
circuit.append(cirq.measure(*qubit, key="Final state"))

# Display the circuit:
printmd("\n**Cirq circuit:**")
print(circuit)

# Invoke the Cirq quantum simulator to execute the circuit:
simulator = cirq.Simulator()

# Simulate the circuit several times:
result = simulator.run(circuit, repetitions=10)

# Print the results:
printmd("\n**Results:**")
print(result)
\end{lstlisting}

The results from the quantum simulator are from 10 measurements in the computational basis of the final state $\ket{\beta_{00}}$, the same ``EPR pair" state as in the Alice and Bob story
told in the next section.
The results for each qubit vary randomly, but are 100\% correlated between the two qubits. Notice that Cirq reports that the final state of both qubits is in the {\it center} of the
Bloch sphere, not on its surface! This is a feature of entanglement that we will come back to in Section \ref{sec:five}.

\begin{verbatim}
Bloch Sphere of the qubit 0 in the final state:

x:  0.0  y:  0.0  z:  0.0

Bloch Sphere of the qubit 1 in the final state:

x:  0.0  y:  0.0  z:  0.0

Cirq circuit:

0: ───H───@───M('Final state')───
          │   │
1: ───────X───M──────────────────

Results:

Final state=1111001000, 1111001000
\end{verbatim}


\subsection{Entanglement and the EPR paradox}\label{ss:EPR}

From the point of view of information science, the most important feature of quantum physics is entanglement. This central role is certainly
not apparent in most of the textbooks familiar to particle theorists, where entanglement is only mentioned in passing as a curiosity, if it is
even mentioned at all. As we shall see however, the study of entanglement overlaps with some fundamental issues of particle physics.

One of the striking features of quantum entanglement is that it does not have any built-in scale. This is in contrast to most quantum effects,
where the scale is set by the dimensionful parameter $\hbar = 6.58$ $\times 10^{-16}$ eV$\cdot$seconds. Because $\hbar$ is tiny, it is
unusual to observe quantum effects on the human scale; I have already mentioned one exception to this: neutrino oscillations, which occur
over scales of a thousand kilometers because of the mysteriously tiny mass differences of the different neutrino flavors. Another exception is ultra-cold atoms,
which are so cold that quantum effects can dominate over the effects of the classical thermal ensemble on scales of meters or more;
this fact will be exploited in the MAGIS-100 experiment currently under construction at Fermilab.

For quantum entanglement, we can plausibly discuss effects on the scales anywhere from the Planck scale of 1.6 $\times 10^{-35}$ meters to the
scale of  light years. Taking the latter extreme, here is a story based on the famous EPR ``paradox" of Einstein, Podolosky, and Rosen.

\begin{quotation}
One day in 2030, a scientist named Carmen uses the DOE quantum internet node at Fermilab to produce $10^9$ pairs of 1550 nanometer wavelength
photons, each pair in the superposition state
\end{quotation}
\begin{eqnarray}
\ket{\beta_{00}} =
\frac{1}{\sqrt{2}}\left( \ket{00} + \ket{11} \right) \nonumber
\end{eqnarray}
\begin{quotation}
\noindent where 0/1 denote two time bins that are 2 nanoseconds apart. Carmen splits each EPR pair of photons and sends them into space in opposite
directions along the galactic disk, one pair of photons every millisecond.

As it happens, 25 light years in either direction from Earth are two inhabited planets, known as Colony A and Colony B of the same advanced
civilization. Just after (using now Galactic Standard Time to synchronize clocks) Carmen sends her signal, a baby named Alice is born on
Colony A, and a baby named Bob is born on Colony B. By the time Carmen's signals arrive, Alice and Bob have grown up and are each
working at gigantic arrays of Superconducting Nanowire Single Photon Detectors (SNSPDs) on their respective planets. Of course on the
day that Carmen's signals arrive, Alice and Bob have never been in causal contact.

Alice and Bob each detect a train of $10^9$ photons, evenly spaced in time except for small shifts that are uniform in size but seem to be
random in pattern. They correctly interpret that these are signals from a primitive civilization on Sol-3, known from spectral observations
to have methane-producing inhabitants.

These discoveries are big news on both planets, and through pecularities common to the two cultures, the sequence of binary bits is
interpreted as a 1 megabyte  header representing a time, and the rest as coordinates on the orthogonal plane containing Sol-3 that bisects a line between the two colonies. 
Both colonies launch space probes on predetermined ballistic trajectories towards the common set of coordinates designed to arrive at the common time. Alice and Bob
are honored to be chosen to ride on the spacecraft of their respective colonies. Alice and Bob and both killed
when the two space probes tragically collide, having arrived at the same coordinates at the same time.

Many years after that, when
communications with Earth are established, the attorneys for Alice and Bob's estates contact Carmen, now a very emeritus professor. They accuse Carmen
of deliberately sending the same sequence of bits to both colonies, with tragic consequences. Carmen tells them that {\it she has never been in possession of this sequence of bits},
and in fact that it would have been {\it physicially impossible for her to have known the sequence} at the time that both colonies launched their spacecraft. The case
is eventually dismissed by the Galactic Supreme Court.
\end{quotation}
 
This long-winded version of the EPR paradox is meant to emphasize certain features:
\begin{itemize}
\item The effects of quantum entanglement can be seen over large distances. As we shall see, entanglement effects have already been demonstrated with
EPR pairs of photons over
distance scales of tens of kilometers. 
\item Quantum randomness in this example occurs between Carmen and Alice, and between Carmen and Bob, but not between Alice and Bob.
\item While Alice and Bob receive the same binary bitstring it is causally impossible at that time for Carmen to be in possession of their bitstring.
\item While randomly generated, the bitstring that they share can eventually have large causal consequences.
\end{itemize}
 
This is the quantum weirdness that Einstein called ``spukhafte Fernwirkung", meaning spooky (or ghostly) action at a distance.
It has been argued by many physicists that this motivates trying to replace quantum mechanics with a more ``reasonable" description
of entangled states using information encoded in some extra ``hidden" variables. However John Stuart Bell's brilliant work has already
allowed direct experimental tests that distinguish between the quantum description and a very general class of hidden variable alternatives.
We will discuss his work in Section \ref{sec:Bellinequality}, after we develop some formalism for measuring entanglement in quantum systems.


\subsection{Experiment 3: A circuit to SWAP two qubit states} 

\begin{figure}[tb]
\centering
\includegraphics[width=0.7\textwidth]{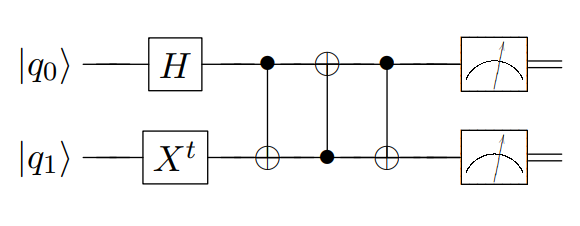}
\caption{A circuit to SWAP two qubit states.The single qubit operation is a Hadamard gate. The two qubit gate
is a CNOT gate. In the circuit shown the initial state is $\ket{00}$, and the two dials indicate measurement
of the final state in the computational basis.\label{fig:Exp3}}
\end{figure}

A quantum circuit for this is illustrated in Figure~\ref{fig:Exp3}. The idea is to prepare two single qubit states - doesn't matter what they are -
then apply a unitary operator that swaps the quantum states of the two qubits. The SWAP procedure can be accomplished using three CNOT gates.
 With the notebook provided you can run this on the quantum simulator using the following Python code:

\begin{lstlisting}[language=Python]
# Get two qubits and a circuit
qubit = [cirq.LineQubit(x) for x in range(2)]
circuit = cirq.Circuit()

# Add a Hadamard gate to make the initial state of qubit 0:
circuit.append([cirq.H(qubit[0])])

# Get a symbol
symbol = Symbol("t")
# Add a parameterized XPowGate to make the initial state of qubit 1:
circuit.append([cirq.XPowGate(exponent=symbol)(qubit[1])])

# Add three CNOT gates to make a SWAP gate:
circuit.append([cirq.CNOT(qubit[0], qubit[1]),
               cirq.CNOT(qubit[1], qubit[0]),
               cirq.CNOT(qubit[0], qubit[1])])

# Measure qubit 1 first, then measure qubit 0:
circuit.append(cirq.measure(qubit[1], key='q1'))
circuit.append(cirq.measure(qubit[0], key='q0'), strategy=InsertStrategy.NEW)

# Display the circuit:
printmd("\n**Cirq circuit:**")
print(circuit)

# Get a sweep over parameter values
sweep = cirq.Linspace(key=symbol.name, start=0.0, stop=1.0, length=3)

# Execute the circuit for all values in the sweep
sim = cirq.Simulator()
results = sim.run_sweep(circuit, sweep, repetitions=50)
printmd("\n**Results for t = 0:**")
print(results[0])
printmd("\n**Results for t = 1:**")
print(results[2])
printmd("\n**Results for t = 0.5:**")
print(results[1])
\end{lstlisting}

In this experiment qubit 0 is always prepared in the state  $\ket{+}$, while for qubit 1 we used three different
choices of initial state according to the value of the parameter $t$ in the ``XPowGate":
\begin{eqnarray}
t=0&:& {\rm XPowGate(t)} = {\rm exp}(i \bm{X}\pi t/2) = \mathbbm{1} \; ; \quad \ket{q_1} = \ket{0}  \\\nonumber
t=1&:& {\rm XPowGate(t)} ={\rm exp}(i \bm{X}\pi t/2) = \bm{iX}  \; ; \quad \ket{q_1} = \ket{1} \\\nonumber
t=0.5&:& {\rm XPowGate(t)} = {\rm exp}(i \bm{X}\pi t/2) =  \frac{1}{\sqrt{2}}\left( \mathbbm{1} + i\, \bm{X} \right)  \; ; \quad \ket{q_1} = \ket{\bm{i}}
\end{eqnarray}
In each of the three cases we run the circuit 50 times, getting the following results:

\begin{verbatim}
Cirq circuit:

0: ───H─────@───X───@─────────────M('q0')───
            │   │   │
1: ───X^t───X───@───X───M('q1')─────────────

Results for t = 0:

q0=00000000000000000000000000000000000000000000000000
q1=01100011111010111111111110101101010100011010111001

Results for t = 1:

q0=11111111111111111111111111111111111111111111111111
q1=10100111011100100101011100101100100001001000101100

Results for t = 0.5:

q0=10100101111000101011000101100101100010100011110111
q1=10111101111001000110010010000011011001000111100111
\end{verbatim}

You see in each case that the two qubits have swapped states, although it is hard to tell the difference in the third case (think about how you could distinguish them).
We will come back to this example when we discuss quantum measurement.


%
%


\subsection{The no-cloning theorem}\label{ss:noclone}

The previous example shows us how quantum information, by which I mean the state information of one or more qubits, can be transferred from one system to another.
However it is easy to see that quantum states cannot be directly copied or cloned via unitary transformations such as gate operations. To be precise, if 
$\ket{a}$ is some particular quantum state and $U$ is some two-qubit unitary transformation such that $U\ket{a\,0} = \ket{a\,a}$, then it is impossible that this property holds for
arbitrary states $\ket{a}$. This follows trivially from the possibility of linear quantum superpositions. Suppose we found two orthogonal states (such as $\ket{0}$ and
$\ket{1}$ in the single qubit case) and a $U$ such that $U\ket{a\,0} = \ket{a\,a}$ and $U\ket{b\,0} = \ket{b\,b}$. Now consider the superposition
$\ket{c} \equiv  (1/\sqrt{2})(\ket{a} + \ket{b})$; obviously:
\begin{eqnarray}
U\ket{c\,0} &=& \frac{1}{\sqrt{2}}\left( U\ket{a\,0} + U\ket{b\,0} \right) =\frac{1}{\sqrt{2}} \left( \ket{a\,a} + \ket{b\,b}\right) \\\nonumber
&\neq & \ket{c\,c} = \frac{1}{2} \left( \ket{a\,a} + \ket{a\,b} + \ket{b\,a} + \ket{b\,b} \right)
\end{eqnarray}

The no-cloning theorem means that we cannot copy an unknown quantum state; we can copy a known quantum state, but this is really no different
from the fact that we can prepare many identical copies of the same known state. The no-cloning theorem has major consequences for quantum
communications, since it implies that you cannot construct an amplifier that preserves arbitrary quantum information.


\section{Quantum teleportation}\label{sec:three}
Quantum teleportation is a process by which a qubit state can be transmitted by sending only two classical bits of information \cite{Bennett:1992tv}.
This is accomplished by pre-sharing a Bell state between the sender (Alice) and the receiver (Bob). 
This entangled state allows the receiver (Bob) of the two classical bits of information to possess a qubit with the same state as the one originally held by the sender (Alice).
In accord with the no-cloning theorem, after performing the teleportation process Alice is no longer in possession of her original qubit state. In this sense the qubit state
has been teleported from Alice to Bob.

\subsection{Experiment 4: A circuit for quantum teleportation} 
In the example provided in the notebook, qubit 0 (the Message) is prepared in a random state by applying X and Y gates. Alice is in possession of both the message qubit and
qubit 1, which is part of an EPR pair with Bob's qubit 2. Alice now performs a Bell state measurement on her pair of qubits, getting one of four possible results. She then
transmits that result (equivalent to two classical bits) to Bob, and Bob performs unitary operations on his qubit that depend on what he received from Alice.
The Cirq quantum circuit is nicely able to handle this.The result is that Bob's qubit final state is guaranteed to be in whatever state the message qubit
was in originally. This is only possible given that an entangled state was pre-shared between Alice and Bob.

%
\begin{figure}[tb]
\centering
\includegraphics[width=0.8\textwidth]{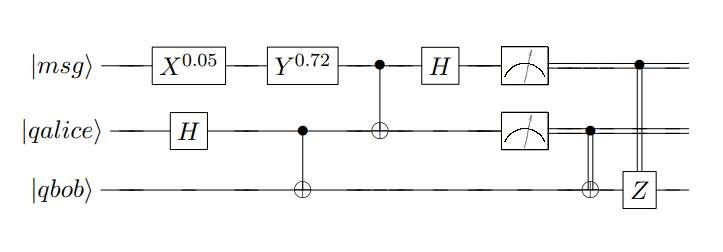}
\caption{Quantum teleportation circuit. In the circuit shown the initial state is $\ket{000}$; the Message qubit is converted to a nontrivial state
by applying random powers of the $\bm{X}$ and $\bm{Y}$ gates. The Alice and Bob qubits are entangled as an EPR pair.
A Bell state measurement is performed on the Message and Alice qubits. Classical information from this measurement then controls
two single qubit operations on the Bob qubit.\label{fig:Exp4}}
\end{figure}
%

\begin{lstlisting}[language=Python]
# Define three qubits: msg = qubit[0], qalice = qubit[1], qbob = qubit[2]
qubit=[0]*(3)
qubit[0] = cirq.NamedQubit('msg')
qubit[1] = cirq.NamedQubit('qalice')
qubit[2] = cirq.NamedQubit('qbob')

circuit = cirq.Circuit()
# Create a Bell state entangled pair to be shared between Alice and Bob.
circuit.append([cirq.H(qubit[1]), cirq.CNOT(qubit[1], qubit[2])])

# Creates a random state for the Message.
ranX = random.random()
ranY = random.random()
circuit.append([cirq.X(qubit[0])**ranX, cirq.Y(qubit[0])**ranY])

# Unitary operator rotating the two-qubit basis of the Message and Alice's entangled qubit;
# rotates the Bell state basis to the computational basis:
circuit.append([cirq.CNOT(qubit[0], qubit[1]), cirq.H(qubit[0])])
# Combining now with a measurment in the computational basis,
# we effectively have projected this two-qubit state onto one of the four states of
# the Bell state basis:
circuit.append(cirq.measure(qubit[0], qubit[1]))

# Use the two classical bits from the Bell measurement to recover the
# original quantum Message on Bob's entangled qubit.
circuit.append([cirq.CNOT(qubit[1], qubit[2]), cirq.CZ(qubit[0], qubit[2])])

printmd("\n**Cirq circuit:**")
print(circuit)

sim = cirq.Simulator()

# Run a simple simulation that applies the random X and Y gates that
# create our message.
q0 = cirq.LineQubit(0)
message = sim.simulate(cirq.Circuit([cirq.X(q0)**ranX, cirq.Y(q0)**ranY]))

printmd("\n**Bloch Sphere of the Message qubit in the initial state:**")
expected = cirq.bloch_vector_from_state_vector(message.final_state,0)
print("x: ", around(expected[0], 4), " y: ", around(expected[1], 4),
      " z: ", around(expected[2], 4))

# Records the final state of the simulation.
final_results = sim.simulate(circuit)

printmd("\n**Bloch Sphere of Bob's qubit in the final state:**")
teleported = cirq.bloch_vector_from_state_vector(
    final_results.final_state, 2)
print("x: ", around(teleported[0], 4), " y: ",
    around(teleported[1], 4), " z: ", around(teleported[2], 4))

printmd("\n**Bloch Sphere of the Message qubit in the final state:**")
message_final = cirq.bloch_vector_from_state_vector(
    final_results.final_state, 0)
print("x: ", around(message_final[0], 4), " y: ",
    around(message_final[1], 4), " z: ", around(message_final[2], 4))
\end{lstlisting}

It is fun to run this several times to see that it works. A characteristic output is:
\begin{verbatim}
Cirq circuit:

msg: ──────X^0.103───Y^0.456───@───H───M───────@───
                               │       │       │
qalice: ───H─────────@─────────X───────M───@───┼───
                     │                     │   │
qbob: ───────────────X─────────────────────X───@───

Bloch Sphere of the Message qubit in the initial state:

x:  0.9396  y:  -0.3169  z:  0.1295

Bloch Sphere of Bob's qubit in the final state:

x:  0.9396  y:  -0.3169  z:  0.1295

Bloch Sphere of the Message qubit in the final state:

x:  0.0  y:  0.0  z:  1.0
\end{verbatim}

Notice that the final state of the Message qubit is always trivial, either $\ket{0}$ or $\ket{1}$; this is a manifestation of the
no-cloning theorem. For fans of Star Trek, this is why Dr. McCoy objected to using the teleportation device on the starship {\it Enterprise}:
the teleportation process appears to destroy the original information and then recreate it instantaneously somewhere else.


%
\begin{figure}[tb]
\centering
\includegraphics[width=0.8\textwidth]{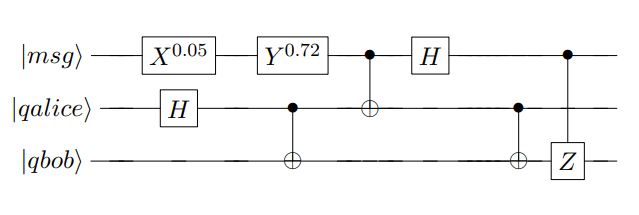}
\caption{Quantum teleportation without measurement. In the circuit shown the initial state is $\ket{000}$; the Message qubit is converted to a nontrivial state
by applying random powers of the $\bm{X}$ and $\bm{Y}$ gates. The Alice and Bob qubits are entangled as an EPR pair.
A unitary basis rotation from the Bell basis to the computational basis is performed on the Message and Alice qubits. These qubits then control
two single qubit operations on the Bob qubit.\label{fig:Exp5}}
\end{figure}
%

\subsection{Experiment 5: Quantum teleportation without measurement}


Your Python notebook contains this interesting variation on the basic quantum teleportation circuit. It is exactly the same as the circuit in Experiment 4,
except that we skip the measurement of qubits 0 and 1 in the computational basis. Note that this is the second step of what I was calling
a Bell state measurement; we still need to do the first step, which was a unitary basis change on qubits 0 and 1.
So all we do is comment out one line of the Cirq code
as follows:

\begin{lstlisting}[language=Python]
# Unitary operator rotating the two-qubit basis of the Message and Alice's entangled qubit;
# rotates the Bell state basis to the computational basis:
circuit.append([cirq.CNOT(qubit[0], qubit[1]), cirq.H(qubit[0])])
# But this time skip the measurement
# circuit.append(cirq.measure(qubit[0], qubit[1]))

# Use the same operations as before to recover the
# original quantum Message on Bob's entangled qubit.
circuit.append([cirq.CNOT(qubit[1], qubit[2]), cirq.CZ(qubit[0], qubit[2])])
\end{lstlisting}

As you can see for yourself by running the circuit, the teleportation works just as well as it did before.
This fact is a manifestation of the {\it principle of deferred measurement}, which implies that the 
operation of measuring a qubit commutes with the operation of using it as a control for a controlled gate operation.
This means that in our quantum teleportation circuit I can delay the measurement until after the unitary controlled gate operations that
act on Bob's qubit - but having delayed those measurements I don't need to do them at all.

The catch here is that now the two controlled gate operations require a direct physical connection between qubits 0 and 2, rendering
teleportation rather less magical. In fact, by comparing with the SWAP circuit of Experiment 3, which also contained three controlled gate
operations, we see that the quantum teleportation circuit without measurement is just a variation on the swap.

The whole magic of teleportation comes from substituting a classical communications channel for part of the swap, and which allows
Bob's qubit to be arbitrarily far away from Alice's qubit. The speed of teleportation is limited only by the fact that the classical information
cannot be transmitted faster than the speed of light. Indeed Bob could choose to measure his qubit {\it before} receiving the classical
transmission from Alice, since 25\% of the time he already has the correct quantum state without performing any ``corrections".
Then when Alice's message finally shows up, Bob will know if his teleported message was valid or not. 

Quantum teleportation over large distances is not a thought experiment anymore. The Alliance for Quantum Technologies INQNET
program, led by Caltech, has already commissioned high performance
quantum teleportation systems at Fermilab and Caltech that use near-infrared photons moving over standard telecom fiber, with two time bins 
as the computational basis for the photonic qubit \cite{Spiropulu:2020}.
Other national labs are developing similar systems, as part of the recently announced DOE Quantum Internet Blueprint \cite{blueprint:2020},
which will connect all 17 DOE national labs with a high performance quantum teleportation network - not quite ``Beam me up Scotty" but still quite impressive.


\section{Information and measurement}\label{sec:four}

This section reviews the basic concepts of classical information theory and of classical measurement.
This language will give us enough knowledge to then discuss in Section \ref{sec:five} the general phenomena
of {\it quantum decoherence}, which in turn will lead to a modern discussion of {\it quantum measurement}.
One of my main goals here is to confront the fundamental
sloppiness of how we all learned (and some of us taught) quantum mechanics in college. 
Here are two standard examples of sloppiness in the way we talk about quantum mechanics:

\begin{itemize}
\item {\bf Claim I:} Classical physics is deterministic, while quantum physics is fundamentally probabilistic.
\item {\bf Claim II:} In classical physics the observer and details of measurement are irrelevant to the state of the system, whereas in quantum physics they play a very special role.
\end{itemize}

By the end of Section \ref{sec:measure}, you will see that both of these claims are (at best)  misleading. Indeed it would be more accurate (though still perhaps misleading) to replace them with the following
statements, which appear to be almost the direct converses:

\begin{itemize}
\item  {\bf Claim I$^\prime$:} Classical physics is fundamentally probabilistic, while quantum physics is determinisitic.
\item  {\bf Claim II$^\prime$:} In classical physics the state of the system depends fundamentally on the observer, whereas in quantum physics states are defined independently of the notion of observers and measurement.
\end{itemize}

A classical computer program takes an input and computes an output. Without loss of generality we
can take the input to be an $n$-bit binary number, and the output to be an $m$-bit binary number. Then a computer
program performs a mapping
\begin{eqnarray} \label{mapping1}
{\rm C}_{n,m}: \{0,1\}^n \to \{0,1\}^m 
\end{eqnarray}
The input bitstring and the output bitstring are examples of information. When we talk about information in our possession, it is customary to measure
it in binary bits. Thus if you own the complete works of William Shakespeare, you have approximately 16,800,000 bits of information.

In classical statistical mechanics we introduce the related concept of entropy, which can be regarded as a measure of information {\it not} in your possession.
Consider Boltzmann's famous equation:
\begin{eqnarray}
S = k_B \,{\rm log}\,\Omega
\end{eqnarray}
where $S$ is the entropy of a physical system specified by the values of some thermodynamic variables, $k_B$ is Boltzmann's constant, and $\Omega$ is
the number of microstates accessible to the system given those fixed values of the thermodynamic variables. 

Claude Shannon, the founder of modern information theory, generalized this notion by defining what is now known as the Shannon entropy for a
general stochastic system:
\begin{eqnarray}
S = -\sum_i p_i \,{\rm log}_2 (p_i)
\end{eqnarray}
where $i$ labels possible configurations of the system, $p_i$ is the probability of that configuration, and from now on when we take logarithms we always mean base 2.


\begin{figure}
  \centering
    \includegraphics[width=0.75\textwidth]{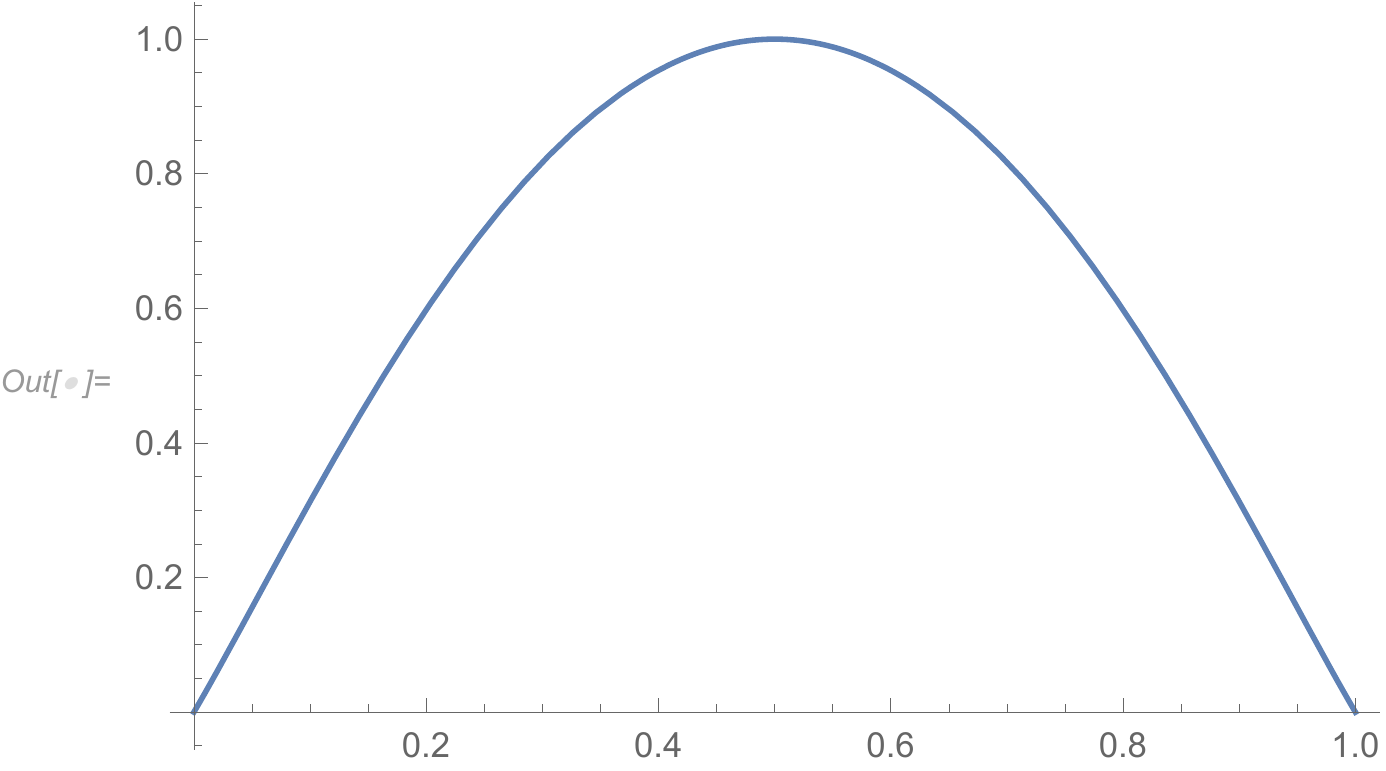}
 \caption{Entanglement entropy for a biased coin flip plotted as a function of the probability that the coin comes up heads.}
\label{fig:coinflip}
\end{figure}


Consider a coin flip. The final state configuration can be described by a single bit, where (say) 0 denotes heads and 1 denotes tails. If the coin flip is unbiased, then there
is a 50-50 chance of either outcome. In this case the Shannon entropy is
\begin{eqnarray}
S = -(\frac{1}{2}\,{\rm log}\frac{1}{2} + \frac{1}{2}\,{\rm log}\frac{1}{2}) = -{\rm log}\frac{1}{2} = 1
\end{eqnarray}
This corresponds to the fact that prior to performing the coin flip we know nothing about the outcome: the possible outcomes are decribed by one bit and our ignorance
(as measured by the Shannon entropy) is also one bit. Note the maximum Shannon entropy, denoting maximum ignorance, is equal to $n$ for a configuration
space labeled by $n$ bits.
 
Now suppose that the coin flip is instead biased, and we know the bias, having estimated it from observing previous coin flips. Then we have partial 
information about the result of the next coin flip, and therefore the Shannon entropy is reduced. Figure \ref{fig:coinflip} shows the Shannon entropy for the coin flip as
a function of the probability that the coin comes up heads.
Note that the Shannon entropy vanishes when the coin flip is completely biased, since in this case we have complete prior knowledge
of the outcome.

Comparing with the Boltzmann entropy, we see that Boltzmann was using the Shannon entropy with the simplifying assumption that all of the 
microstates are equally likely.

\subsection{Classical measurement theory}\label{ss:cmt}

In classical physics measurements are performed on a configuration space of system variables. Examples of system variables are position, momentum, temperature, pressure,
height, color,...whatever you like.
It is up to the observer to specify the configuration space of
interest, and the observer may already has some partial information about the system. This prior information in general is a probability distribution, so we can
define the classical prior state as a probability distribution on an observer-specified configuration space. For example, suppose I am interested in where in the
world is Carmen Sandiego. I might decide to simplify my configuration space to a single bit, where 0 denotes that Carmen is currently in the northern hemisphere, and
1 that she is in the southern hemisphere. You might decide to use a different configuration space, perhaps with 0/1 denoting that Carmen is in the western/eastern
hemisphere. Even if you chose the same definition of the configuration space, you might have different partial information than I have about Carmen's current location. In both
cases you and I are constructing {\it different} classical prior states to describe Carmen; the classical state description is tied fundamentally to the observer. There is nothing
strange about this as long as different observers construct {\it consistent} descriptions of the same physical system.

A classical measurement is a process for collecting more information about a particular classical system. The measurement itself consists of some readout variables
defining the configuration space of the measuring apparatus,
that we denote collectively by $Y$, which are supposed to be correlated with, and thus give us information about, the system variables $X$.
A classical measurement of continuous system variables has no pre-defined limit on its precision, and thus we can imagine measuring position
and momentum simultaneously with a precision that would cause Heisenberg to object. However any actual classical measurement has limited precision,
since your readout gauges have a finite number of registers; thus we can always think about classical measurement in terms of discrete variables.

Classical measurements in general are imperfect, due to the presence of noise. This is another way of saying that the combined configuration space
of your system+apparatus is always ignoring many other degrees of freedom, introducing unknown dynamics that can decorrelate the readout
variables from the system variables. In order to interpret the classical measurement we need to introduce some kind of modeling of the noise,
often in terms of some random variables that I denote by $\Xi$. In this notation the values obtained for the readout variable in the
classical measurement are some function of the system variables and the noise variables:
\begin{eqnarray}
Y = F(X,\Xi )
\label{eq:map}
\end{eqnarray}
Since we can regard $X$ and $Y$ as binary bitstrings (of equal length), a simple example of a noise model is to introduce the possibility of
{\it bit-flip errors} in the readout:
\begin{eqnarray}
Y = X \oplus \Xi
\label{eq:bitflip}
\end{eqnarray}
where $\oplus$ is the Boolean ``exclusive or" operator, also known as XOR, which takes $00\to 0$, $01\to 1$, $10\to1$, and $11\to 0$.

In the simple case of a single bit system $x=0,1$, we assume that the noise variable is also a single bit $\xi=0,1$, and that $x$ and $\xi$ combine
via the XOR operator
to determine the single bit readout variable $y=0,1$.
The simple bit-flip noise model assumes that there
is a probability $1-\mu$ that the readout bit ($y=0,1$) has the bit-flipped, i.e. wrong, value. Thus ${\cal P}(\xi = 0) = \mu$, 
${\cal P}(\xi = 1) = 1-\mu$, and (\ref{eq:bitflip}) can be written:
\begin{eqnarray}
{\cal P}(y=0) &=& \mu\,{\cal P}(x=0) + (1-\mu)\,{\cal P}(x=1) \\\nonumber
{\cal P}(y=1) &=& (1-\mu)\,{\cal P}(x=0) + \mu\,{\cal P}(x=1)
\label{eq:py}
\end{eqnarray}

In this simple example our classical prior state is completely specified by ${\cal P}(x=0)$, since ${\cal P}(x=1) = 1 -{\cal P}(x=0)$. After we perform
the classical measurement and obtain a value for the readout variable $y$, we use this information to define the post-measurement system state by
${\cal P}(x|y)$, where this notation means ``the probability of the system variable having value $x$ given that the readout variable had value $y$''.
This {\it posterior} classical system state is computed using the famous Bayes' theorem:
\begin{eqnarray}
{\cal P}(x|y) = \frac{{\cal P}(y|x) * {\cal P}(x)}{{\cal P}(y)}
\end{eqnarray}

Using Bayes' theorem it is straightforward \cite{Wiseman} to derive a formula that computes ${\cal P}(x|y)$ given any prior ${\cal P}(x)$, our noise model, and the result $y$ of the measurement.
If the result
of the measurement is $y=0$ then the posterior state is
\begin{eqnarray}
{\cal P}(x|y) = \frac{{\cal P}(\xi = x)*{\cal P}(x)}{{\cal P}(\xi = 0){\cal P}(x=0) + {\cal P}(\xi = 1){\cal P}(x=1)}
\end{eqnarray}
whereas if the measurement gives $y=1$ the posterior state is
 \begin{eqnarray}
{\cal P}(x|y) = \frac{{\cal P}(\xi = !x)*{\cal P}(x)}{{\cal P}(\xi = 0){\cal P}(x=0) + {\cal P}(\xi = 1){\cal P}(x=1)}
\end{eqnarray}
where $!x$ means NOT $x$.

From this result you can check that,
in the limit of a zero noise measurement, meaning ${\cal P}(\xi = 0) = \mu \to 1$, you get ${\cal P}(x|y) \to \delta_{x,y}$, meaning that
the posterior state now has complete information about whether Carmen is currently in the northern or southern hemisphere.
In the limit of maximally bad noise,  ${\cal P}(\xi = 0) = \mu \to 1/2$, you can check that  ${\cal P}(x|y) \to {\cal P}(x)$, meaning that
the measurement {\it did not add any new information.}

\subsection{Review of classical claims, with help from the LHC}

Let us now re-examine the parts of Claims I,II and I$^\prime$,II$^\prime$ that referred to classical systems. The assertion in Claim II$^\prime$ that
in classical physics the state of the system depends fundamentally on the observer is true by definition, as we have just reviewed.
The assertion in Claim I$^\prime$ that classical physics is fundamentally probabilistic is also true by definition, as we have seen.
The strategy of the classical physicist is to choose a configuration space with some manageable number of
system variables to describe part of a classical system and part of its dynamics, then perform measurements that increase our information
about the system variables. In particular the classical physicist looks for predictive regularities which an old school physicist would codify as ``laws", 
or in a more modern approach treat as correlations of the system variables. In the modern language we say that the physical world
has some underlying ``latent variables" $Z$ with correlations between them, $P(Z)$. When we go out and make measurements of some
observer-defined  set of system variable $X$, we are observing an instance of $P(X,Z)$, the joint probability distribution that relates 
some underlying latent variables to the measurable  system variables. Our job as physicists is to use a few of these instances as clues
to guess the underlying ``generative model"; for example by making measurements of the orbits of a few planets I infer a generative model
that explains all possible orbits of all possible planets. Just as in machine learning, the power of this procedure is that often a generative model
with a small number of latent variables $Z$ does an excellent job of generating new instances of some particular set of measurable system
variables $X$ with features that we happen to be interested in; this is what a physicist calls ``making predictions". Just as in machine learning,
this works best when the features that we care about are coarse-grained ``macro" features.
Probabilities are fundamental to classical descriptions
because in the coarse-graining that we use to define a particular macro configuration space we treat other ``micro" states as statistical
ensembles. 

These facts are well known to experimental high energy particle physicists in the modern era, who analyze data from
particle detectors. Indeed I first learned about Bayes' theorem from my CMS colleague Maurizio Pierini, one of the world's leading
experts on how to extract the maximum amount of information from collider datasets. An analysis of CMS data describes the end result of a
proton-proton collision in the CMS detector by a finite number of classical system variables $X$, such as four-momenta and charges of final state particles,
mapped to a set of discrete readout variables $Y$, which could be hits in a tracker layer or energy deposits in a calorimeter cell,
with a mapping of the form (\ref{eq:map}). The analog
of $\Xi$ and the function $F(X,\Xi)$ are models and probability distributions (obtained ultimately from data) of various aspects of detector response.
Although some quantum effects creep in, these probabilities are best regarded as classical in nature: 
they arise because knowing that a 100 GeV pion started a shower in the CMS hadronic calorimeter
does not give you enough information to determine the response of the calorimeter, which depends on details of that particular shower.
The choice of the system variables and thus of the classical configuration space for the analysis depends on what the
analyzer is looking for.

What now are we to make of the ``sloppy" statement in Claim I, that classical physics is deterministic.
Classical measurements in a suitably restricted configuration space can be both highly reliable and predictive; for example I am pretty
sure that I am currently located in the northern hemisphere, and can predict with high confidence that I will still be in the
northern hemisphere tomorrow. These sorts of deterministic statements, however, depend very much on your choice of
configuration space. For example, suppose I notice a speck of dust in my hair, and I want to predict
in which part of my house that speck will be tomorrow. This is certainly a classical prediction, but any reasonable classical
physicist would make such a prediction in terms of a probability distribution. We often assert that ``in principle" you can
expand your classical configuration space and your set of classical measurements in such a way as to make deterministic 
predictions of whatever we like; but what we actually mean by this assertion is that in classical physics there are no built-in
hard limits on this process. Similar reasoning applies to the sloppy statement in Claim II that
in classical physics the observer and details of measurement are irrelevant to the state of the system. What this
really means is that we can often choose the classical configuration space and the classical measurement apparatus in such
a way that there is essentially zero chance of back-reaction from the measurement on the chosen system variables.
Thus it is highly unlikely that in the course of measuring that I am currently in the northern hemisphere I will somehow
end up in the southern hemisphere. However, suppose my single bit configuration space is whether or not there are
gasoline fumes in a metal pail, and my classical measurement consists of throwing a lighted match into the pail. In this
case the posterior state is always 0 regardless of whether the prior state was 0 or 1. This is still a perfectly valid classical
measurement.

\section{Quantum entropy and quantum decoherence}\label{sec:five}

\subsection{The density matrix and quantum entropy}
The {\it density matrix} in quantum mechanics is a generalization of the notion of a quantum state vector. It gets around what is otherwise a fundamental
issue of how to describe an {\it open} system, i.e. a quantum system that interacts with degrees of freedom outside the Hilbert space that we are considering (or have
access to). As we have seen, in classical descriptions we describe open systems by a classical ensemble: a sum over the states of the configuration space of interest,
weighted by probabilities determined by interactions with the larger system. This approach will fail spectacularly for open quantum systems, since in general there
will be entanglement between the Hilbert space of interest and the larger system. 

The density matrix solves this problem by defining what looks like a classical ensemble of projection operators onto some $N$-dimensional orthonormal basis of quantum states
in the Hilbert space of interest for some open quantum system; this {\it density operator}  is written as
\begin{eqnarray}
\rho = \sum_{i,j} \, p_{ij} \,\ket{\psi_i} \bra{\psi_j}
\label{eq:rhodef}
\end{eqnarray}
where $\ket{\psi_i}$ are the orthonormal basis states. The matrix elements of the operator $\rho$ are just
the weights $p_{ij}$; these define a basis-dependent $N\times N$ matrix $\rho_{ij}$ called
the density matrix. We require that the density matrix of any quantum open system satisfies three basis-independent properties:
\begin{itemize}
\item $\rho$ is Hermitian.
\item $\bm{tr}\rho = 1$.
\item The eigenvalues of $\rho$ are non-negative.
\end{itemize}

As we noted in Section (\ref{sec:two}), for any quantum state vector written as a superposition, $\ket{\tilde{\psi}} = \sum_i \alpha_i \ket{\psi_i}$, we can always find
a new orthonormal basis where the same state is just one of the basis states. Comparing to Eq. (\ref{eq:rhodef}), this means that the state vector corresponds,
in a suitable basis, to a diagonal density matrix with a single entry equal to 1, and the rest zeroes. You can easily check yourself that not only is $\bm{tr}\rho =1 $ in this case,
but also $\bm{tr}(\rho^2) = 1$. Any density matrix that satisfies  $\bm{tr}(\rho^2) = 1$ is called a {\it pure state}.
Obviously the quantum mechanics of pure states is equivalent to the usual description of a {\it closed} quantum system in terms of state vectors, so
nothing has been gained so far, although the density matrix version of a state vector has the nice property that any overall phase in the state vector
automatically drops out from $\rho$.

Now consider an open quantum system $A$ with Hilbert space $H^A$ that has physical interactions, possibly including entanglement, with another quantum system
$B$ having Hilbert space $H^B$. Physics in the larger space $H^A \otimes H^B$ can be described by pure states corresponding to density matrices $\rho_{AB}$
that satisfy  $\bm{tr}\rho_{AB}^2 = 1$. In order to get a description of physics in the subspace $H^A$ that does not make a reference to what is going on in
$H^B$, we can do a {\it partial trace} of the density operator $\rho_{AB}$ over the subspace $H^B$. The resulting density matrix $\rho_A$ will then describe
the open quantum system with Hilbert space $H^A$.

This is easiest to understand by looking at a simple example. Suppose that Alice and Bob share the EPR pair  $(1/\sqrt{2})( \ket{00} + \ket{11} )$. In this
example $H^A$  and $H^B$ are both single qubit Hilbert spaces, and $H^{AB}$ has dimension $2^2 = 4$. In the standard computational basis for
$H^{AB}$, the EPR pure state corresponds to the following density operator:
\begin{eqnarray}
\rho_{AB} &=& \left[ \frac{1}{\sqrt{2}}\left( \ket{00} + \ket{11} \right) \right]\left[  \frac{1}{\sqrt{2}}\left( \bra{00} + \bra{11} \right) \right] \\
&=& \frac{1}{2} \left( \ket{00}\bra{00} + \ket{00}\bra{11} + \ket{11}\bra{00} + \ket{11}\bra{11} \right) \label{eq:rhoAB}\\
&=& \frac{1}{2}
\begin{pmatrix}
1 & 0 & 0 & 1 \\
0 & 0 & 0 & 0 \\
0 & 0 & 0 & 0 \\
1 & 0 & 0 & 1 
\end{pmatrix}
\end{eqnarray}
Now we can take the partial trace over $H^B$ by summing over the expectation values of the density operator as written in Eq. (\ref{eq:rhoAB}) between the 
computational basis states of $H^B$:
\begin{eqnarray}
\rho_A \equiv  \bm{tr}_B(\rho_{AB}) &=& \bra{0_B} \rho_{AB} \ket{0_B} +  \bra{1_B} \rho_{AB} \ket{1_B} \\
&=& \frac{1}{2} \left( \bra{0_B}  \ket{0_A0_B}\bra{0_A0_B} \ket{0_B} +  \bra{1_B}  \ket{1_A1_B}\bra{1_A1_B} \ket{1_B} \right) \\
&=& \frac{1}{2} \left( \ket{0_A}\bra{0_A} + \ket{1_A}\bra{1_A} \right) \\
&=& \frac{1}{2} 
\begin{pmatrix}
1 & 0 \\
0 & 1 
\end{pmatrix}
\label{eq:Alicefinal}
\end{eqnarray}
The resulting matrix $\rho_A$ is sometimes called the {\it state matrix} of the open quantum system $A$, since it is the appropriate
quantum generalization of the state vector of a closed quantum system. {\it This is the fundamental way to describe an open quantum system.}
Observe that $\rho_A$ is not equivalent to any quantum state vector, since it is not a pure state; we can confirm this by computing the trace
of its square:
\begin{eqnarray}
\bm{tr}(\rho_A^2) = \bm{tr}\left( 
\frac{1}{2} 
\begin{pmatrix}
1 & 0 \\
0 & 1 
\end{pmatrix}
\frac{1}{2} 
\begin{pmatrix}
1 & 0 \\
0 & 1 
\end{pmatrix}
\right)
= \frac{1}{4}\bm{tr}
\begin{pmatrix}
1 & 0 \\
0 & 1 
\end{pmatrix}
= \frac{1}{2} \label{eq:rhoA}
\end{eqnarray}

Density matrices with the property $\bm{tr}(\rho^2) < 1$ are called {\it mixed states}. By using a basis $\ket{\psi_i}$ that diagonalizes such a density matrix, we
can always write it in the form:
\begin{eqnarray}
\rho_{\rm mixed} = \sum_i \,p_i \ket{\psi_i}\bra{\psi_i}
\end{eqnarray}
which looks like a classical ensemble of pure states with probability weights $p_i$. Thus in our simple example of Alice's single qubit, 
the density matrix in Eq. (\ref{eq:Alicefinal}) is a single qubit mixed state. It could indeed describe a classical ensemble where there
is a 50\% chance that we are in the state $\ket{0}$ and a 50\% chance that we are in the state $\ket{1}$. But in fact in our case
$\rho_A$ is the partial trace over an entangled two-qubit state, and the weights $p_i$ are a purely quantum effect: they represent
a loss of {\it quantum information} about the two-qubit entangled state. 

The standard measure of the loss of quantum information from entanglement is called the {\it Von Neumann entropy} or {\it entanglement entropy}:
\begin{eqnarray}
S(\rho) \equiv -\bm{tr}\,\rho\,{\rm log}\,\rho
\end{eqnarray}
For a closed quantum system the entanglement entropy vanishes. For an open quantum system $A$ entangled with another system $B$,
we can compute the entanglement entropy as
\begin{eqnarray}
S_A =  -\bm{tr}\,\rho_A\,{\rm log}\,\rho_A = -\bm{tr_A}\,\left(\bm{tr_B }(\rho_{AB}) {\rm log}\,\bm{tr_B }(\rho_{AB})  \right)
\end{eqnarray}

We see also from using a basis where $\rho_{AB}$ is diagonal that $S_A = S_B$. In the case of Alice $S_A = 1$, representing the fact
that because of the maximal entanglement, Alice is missing a full qubit worth of quantum information. Since $S_B = 1$, Bob is doing no better,
also missing a full qubit worth of quantum information. Another way of saying the same thing is to define the {\it mutual information} $MI$:
\begin{eqnarray}
MI(A,B) \equiv S_A + S_B - S_{AB}
\end{eqnarray}
In our simple example, $S_{AB} = 0$ since it comes from a pure state, so $MI = 1 + 1 - 0 = 2$. This represents the maximal amount of 
quantum information that two qubits can share through entanglement.

For a single qubit mixed state, the entanglement entropy has a nice representation in terms of the Bloch sphere. A single qubit density
matrix is a a $2\times 2$ Hermitian matrix with unit trace; this can be described in terms of three real parameters in the basis of the
Pauli matrices:
\begin{eqnarray}
\rho_1 = \frac{1}{2} \left( \mathbbm{1} + x\sigma^x + y\sigma^y + z\sigma^z \right)
\end{eqnarray}
The determinant of this matrix is given by:
\begin{eqnarray}
{\rm det}(\rho ) = \frac{1}{4}(1-r^2)\; ; \quad r = \sqrt{x^2 + y^2 +z^2}
\end{eqnarray}
From the required properties of the density operator, this determinant is the product of two nonnegative eigenvalues, both $\leq 1$.
So $0 \leq r \leq 1$. Furthermore $r=1$ implies that one of the eigenvalues is 1, which means that the other eigenvalue vanishes, since
$\rho_1$ has unit trace. Thus $r=1$ means we have a pure state. Putting this all together, a general single qubit density matrix maps to
a point {\it in the interior of the Bloch sphere}, reaching the surface of the Bloch sphere only for pure states, i.e. states that map to the
single qubit state vector Eq. (\ref{eq:qubit}).

You can check that the entanglement entropy of a general single qubit system can be written as
\begin{eqnarray}
S(\rho_1) = -\left( \left( \frac{1+r}{2} \right) {\rm log}\,  \left( \frac{1+r}{2} \right) +  \left( \frac{1-r}{2} \right) {\rm log}\,  \left( \frac{1-r}{2} \right) \right)
\end{eqnarray}

In our simple example, Alice's qubit, considered as a closed system $A$, is at the exact center of the Bloch sphere, $r=0$. This corresponds to the fact
that, treating $A$ as a closed system, the density matrix $\rho_A$ of Eq. (\ref{eq:rhoA}) does not correspond to any pure single qubit state, 
rather {\it it is the density matrix of a classical system} with
50\% probability of being in state $\ket{0}$ and 50\% probability of being in state $\ket{1}$. Thus Alices's qubit is part of a quantum pure state when
considering the full system $A + B$, but looks like a ``post-measurement" classical state when considering $A$ as a closed system. Of course something like
this has to be true, since as we saw in our example Alice has no causal way of knowing if the pure state of $A+B$ has been measured by Bob or not.  
We will return to this connection between entanglement and quantum measurement in the next section.

\subsection{Quantum decoherence}\label{ss:decohere}

In the real world all quantum systems are open systems; they have some interaction with their environment, an environment that usually consists of a  large number
of unobserved degrees of freedom. At the quantum level these interactions produce entanglement between the ``System"  $S$ that we are trying to model (or manipulate), 
and the ``Environment" $E$. As a result there is a tendency for quantum information in $S$ to be lost, i.e. for growth of the entanglement entropy $S(\rho_S)$ relative to the
combined system $S+E$. I will use the phrase ``quantum decoherence" to denote this environmentally induced process. Quantum decoherence of qubit states in quantum computers
is one of the biggest challenges limiting the performance of today's quantum computers. There is a fundamental tension that afflicts any conceivable hardware
implementation of a digital quantum processor: on the one hand you need to be able to manipulate your qubits in order to create a quantum circuit and run it
and read out the results, but on the
other hand you want to isolate your qubits from outside influences to avoid quantum decoherence.

Of course, as always in quantum physics, the underlying interactions involve only unitary time evolution and are in that sense completely deterministic as viewed in $S+E$, but only in very simplified cases do we have the ability to model or observe
$S+E$. The traditional attitude of the particle physicist is to ignore the existence of the environment entirely, or (even worse) to model the environmental effects entirely
as classical noise, similar to our discussion in Section \ref{sec:four}.

Let's try to understand quantum decoherence in more detail by taking $S$ to consist of a single qubit. At time $t=0$ we manage to prepare $S$ in some particular known
pure state; this means that there exists a basis where
\begin{eqnarray}
\rho_S = 
\begin{pmatrix} 1 & 0 \\ 0 & 0 \end{pmatrix}
\;;\quad \bm{tr}(\rho_S) = 1,
\quad  \bm{tr}(\rho^2_S) = 1
\end{eqnarray}
Interaction with the environment means that there is a larger density matrix $\rho_{SE}$ such that, at any time $t$, $\rho_S$ is given by the partial trace
of $\rho_{SE}$ over a basis in the Hilbert space $H_E$ of the relevant part of the environment. Quantum decoherence is then the tendency for
$S(\rho_S)$ to increase over time, or equivalently for $\bm{tr}(\rho^2_S)$ to decrease over time. For our single qubit example maximal decoherence occurs
when $S(\rho_S) \to 1$, or equivalently $\bm{tr}(\rho^2_S) \to 1/2$. When this happens, regardless of where on the surface of the Bloch sphere we had arranged
to put our qubit, it has now migrated to the center of the sphere.

A key property of the decoherence process is that {\it  interactions with the environment typically select a preferred basis}. This is obvious in the case where the interactions
between $S$ and $E$ are due to an interaction Hamiltonian $H_{\rm int}$ that decomposes as a Hermitian tensor product operator in $S\otimes E$; in such a case
there are observables defined in $S$ that commute with $H_{\rm int}$, and thus there are states in $S$ unaffected by the environmental interaction.
This environmentally-preferred basis is called a {\it pointer basis}, for reasons which will become apparent when
we discuss quantum measurement.
For a single
qubit state the intelligent quantum computer designer will define the computational basis to be this preferred basis; this at least ensures that the states
$\ket{0}$ and $\ket{1}$ are relatively stable.

Let's consider a very simple example. Suppose my single qubit corresponds to a particular transition in an atom: $\ket{0}$ is the lower energy state of the
transition, and $\ket{1}$ is the higher energy state. The environmental interaction that we will consider are photons with frequency close to the atomic transition. 
Up to an overall constant, we can model the interaction by the {\it Jaynes-Cummings} Hamiltonian \cite{NC}:
\begin{eqnarray}
H_{\rm int} =-\left( c_1\,\sigma^z + c_2\,(a^\dagger\,\sigma^- + a\,\sigma^+ )\right)\; ;\quad \sigma^+ = \sigp\; ;\quad \sigma^- = \sigm
\label{eq:HJC}
\end{eqnarray}
where the Pauli matrices act on the atomic qubit, $a^\dagger$,$a$ are creation/annihilation operators for a photon, and $c_1$, $c_2$ are constants.
To make things even simpler, we will only allow single photons, so the photon can also be represented by a single qubit. We will prepare the
initial state at $t=0$ to be $\ket{\psi(t=0)} = \ket{01}$, i.e. the atom in the $\ket{0}$ state along with an occupied single photon state $\ket{1}$.
The density matrix for the two-qubit system at $t=0$ can be written as
\begin{eqnarray}
\rho_{SE}(0) = P^0_1 \otimes P^1_2 \;;\quad P^0_{1,2} = \ket{0}\bra{0} = \proja \;,\quad P^1_{1,2} = \ket{1}\bra{1} = \projb
\end{eqnarray}
where $P^0$, $P^1$ are the single qubit projection operators onto the states $\ket{0}$,$\ket{1}$.

The two-qubit system has a unitary time evolution
\begin{eqnarray}
\ket{\psi(t)} = U(t)\ket{\psi(t=0)} = {\rm e}^{-iH_{\rm int}t}\;\ket{01}
\end{eqnarray}
Using $\rho_{SE}(t) = U(t)\rho_{SE}(0)U^{-1}(t)$ and taking the partial trace, you should be able to derive that
\begin{eqnarray}
\rho_S(t) =\frac{1}{2}\begin{pmatrix}
1+{\rm cos}^2(\sqrt{2}t) & 1 \\ 
0 & {\rm sin}^2(\sqrt{2}t)
\end{pmatrix}
\end{eqnarray}
In the derivation you can use the fact that for a $2^2\times 2^2$ matrix of the form
\begin{eqnarray}
\begin{pmatrix}
a1 &  a2 & a3 & a4 \\
b1 & b2 & b3 & b4 \\
c1 & c2 & c3 & c4 \\
d1 &  d2 & d3 & d4
\end{pmatrix}
\end{eqnarray}
the partial trace over the second qubit gives the $2\times 2$ matrix
\begin{eqnarray}
\begin{pmatrix}
a1 + b2 & a3 + b4 \\
c1 + d2 & c3 + d4
\end{pmatrix}
\label{eq:partial}
\end{eqnarray}

\begin{figure}
  \centering
    \includegraphics[width=0.75\textwidth]{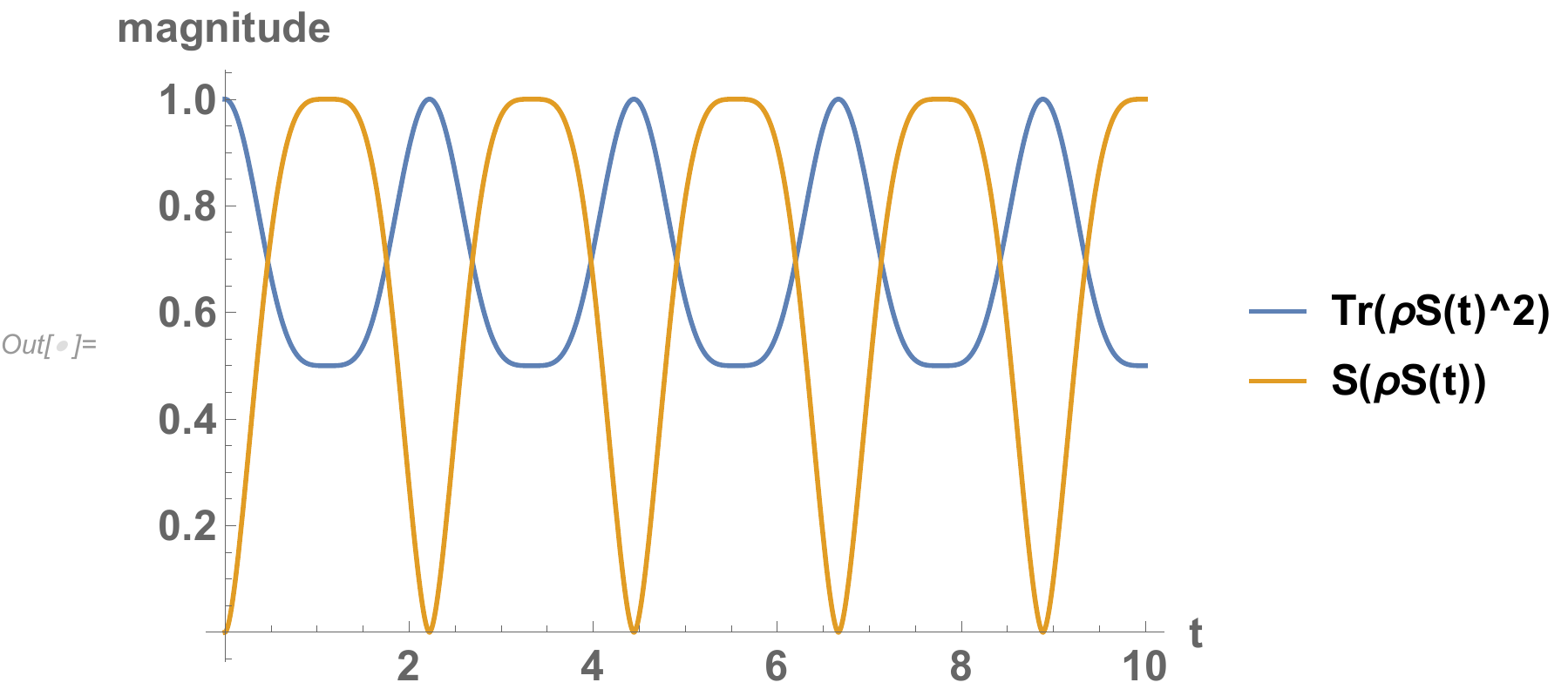}
 \caption{The trace of the square of the single qubit ``System" density matrix $\rho_S(t)$ and the corresponding entanglement entropy,
 plotted versus time after coupling to the ``Environment" qubit with a Jaynes-Cummings type Hamiltonian Eq. (\ref{eq:HJC}) with $c_1 = c_2 = 1$.}
\label{fig:Rabi}
\end{figure}

In Figure (\ref{fig:Rabi}) we see the  trace of the square of the single qubit ``System" density matrix $\rho_S(t)$ and the corresponding entanglement entropy $S(\rho_s(t))$,
plotted versus time. The oscillations are a slightly nonstandard variation of the famous {\it Rabi oscillations} which are used in real quantum hardware to manipulate
qubit states \cite{NC}.
In our case we have cooked things up such that the two-qubit system oscillates between the initial state $\ket{01}$ and the Bell state 
$\ket{\beta_{01}} = (1\sqrt{2})(\ket{01} + \ket{10}$. We have rather arbitraily chosen the basis that diagonalizes the Pauli matrix $\sigma^z$ that appears in $H_{\rm int}$ as the pointer basis; in the usual
demonstrations of Rabi oscillations this is the right choice since $c_1$ is much larger than $c_2$; note that for the underlying physical system the pointer basis is the
energy eigenstate basis, as is often the case.

We see that the interaction with this very simple environment does indeed lead to nonzero entanglement entropy, which reaches its maximum value of one
whenever $t$ is a multiple of $\sqrt{2}\pi/4$, corresponding to a Bell state in $S+E$, and a state at the center of the Bloch sphere in $S$.
We have not achieved quantum decoherence however, because of the simple oscillatory behavior. This is not surprising since we expect true
quantum decoherence to require interacting with (directly or indirectly) a large number of degrees of freedom.

Somewhat surprisingly, we can get pretty close to the true behavior of quantum decoherence by adding one additional environmental qubit.
In analogy with Eq. (\ref{eq:HJC}) I will use the following 3-qubit Hamiltonian:
\begin{eqnarray}
H_{\rm int} &=& -\Bigl( 
c_{11}(\sigma^z_1\otimes  \mathbbm{1}_2\otimes  \mathbbm{1}_3) 
+c_{22}( \mathbbm{1}_1 \otimes \sigma^z_2 \otimes  \mathbbm{1}_3) 
+c_{33}(  \mathbbm{1}_1\otimes  \mathbbm{1}_2 \otimes \sigma^z_3 ) \\ \nonumber
&+& c_{12}(\sigma^+_1\sigma^-_2 + \sigma^-_1\sigma^+_2)\otimes  \mathbbm{1}_3 
+  c_{13}(\sigma^+_1\sigma^-_3 + \sigma^-_1\sigma^+_3)\otimes  \mathbbm{1}_2 
 \Bigr)
\label{eq:HJC3}
\end{eqnarray}

We choose the couplings as follows: $c_{11} = 0.9$, $c_{22} = 0.3$, $c_{33} = 0.4$, $c_{12}=0.5$, $c_{13} = 0.4$. As in the previous example, we
are assuming that the pointer basis is the basis that diagonalizes the Pauli $\sigma^z$ operators, i.e. the usual computational basis.

For the initial state
at $t=0$ we will put our System qubit in the Hadamard state $\ket{+} = (1\sqrt{2})(\ket{0} + \ket{1})$, which is a superposition state in the pointer basis with
\begin{eqnarray}
\rho_S(0) = \frac{1}{2}\begin{pmatrix} 1 & 1 \\ 1 & 1 \end{pmatrix}
\end{eqnarray}
For the two Environmental qubits we will assume they are in the state $\ket{01}$.
As in the previous example we construct the unitary time evolution for the full system $S+E$:
\begin{eqnarray}
\ket{\psi(t)} = U(t)\ket{\psi(t=0)} = {\rm e}^{-iH_{\rm int}t}\;\ket{01}
\end{eqnarray}
We then use $\rho_{SE}(t) = U(t)\rho_{SE}(0)U^{-1}(t)$ and take the partial trace to get $\rho_S(t)$. This example is too difficult to do by hand,
but is easy to implement in {\it Mathematica} or Python. For the partial trace one needs to know that for a $2^3 \times 2^3$ matrix of the form
\begin{eqnarray}
\begin{pmatrix}
 a1 &  a2 &  a3 &  a4 &  a5 &  a6 &  a7 &  a8 \\
 b1 &  b2 &  b3 &  b4 &  b5 &  b6 &  b7 &  b8 \\
 c1 &  c2 &  c3 &  c4 &  c5 &  c6 &  c7 &  c8 \\
 d1 &  d2 &  d3 &  d4 &  d5 &  d6 &  d7 &  d8 \\
 e1 &  e2 &  e3 &  e4 &  e5 &  e6 &  e7 &  e8 \\
 f1 &  f2 &  f3 &  f4 &  f5 &  f6 &  f7 &  f8 \\
 g1 &  g2 &  g3 &  g4 &  g5 &  g6 &  g7 &  g8 \\
 h1 &  h2 &  h3 &  h4 &  h5 &  h6 &  h7 &  h8 
\end{pmatrix}
\end{eqnarray}
the partial trace over the second and third qubits gives 
the $2\times 2$ matrix
\begin{eqnarray}
\begin{pmatrix}
 a1 + b2 + c3 + d4 &  a5 + b6 + c7 + d8 \\
 e1 + f2 + g3 + h4 &  e5 + f6 + g7 + h8
\end{pmatrix}
\end{eqnarray}

\begin{figure}
  \centering
    \includegraphics[width=0.75\textwidth]{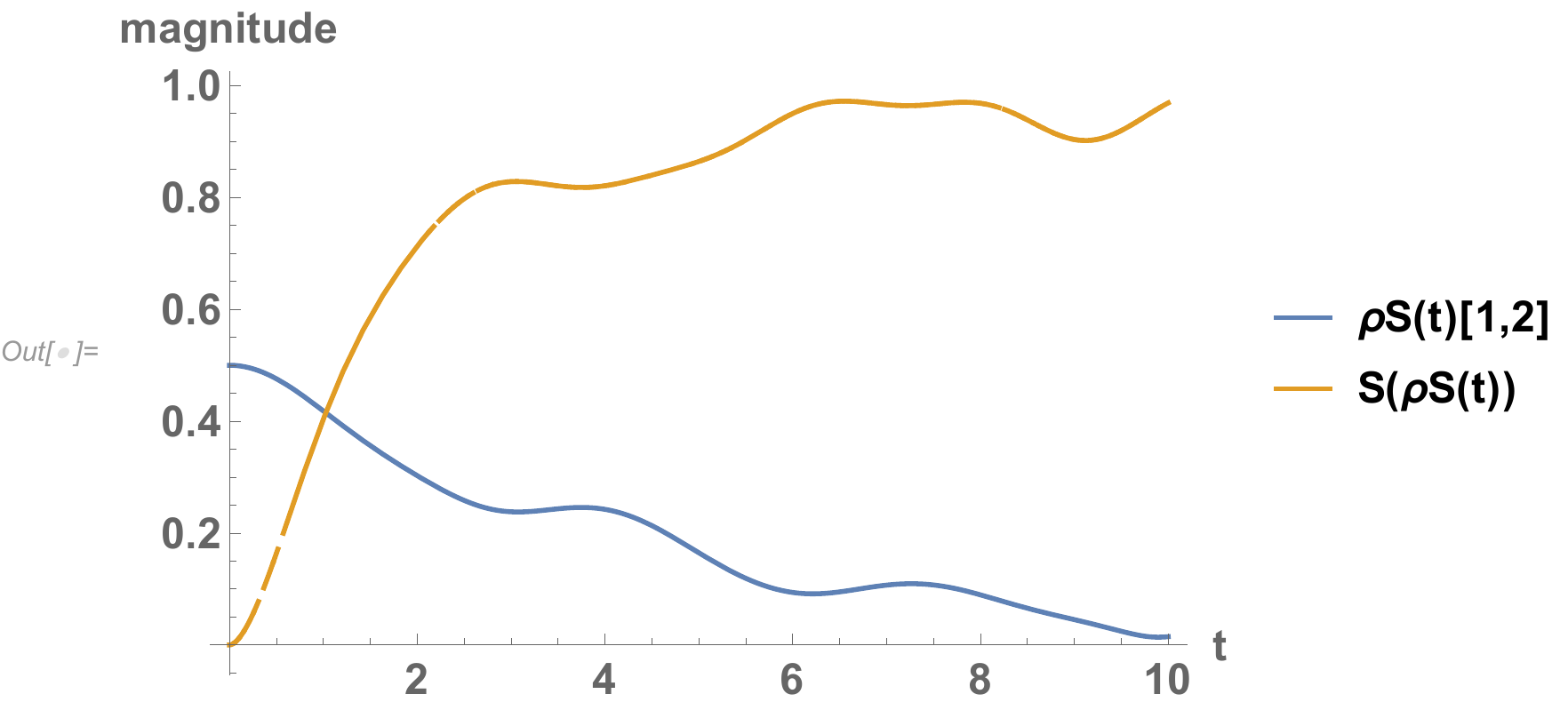}
 \caption{The absolute value of the off-diagonal term in the single qubit ``System" density matrix $\rho_S(t)$ and the corresponding entanglement entropy $S(\rho_S(t))$,
 plotted versus time after coupling to the two ``Environment" qubits. The initial System state is the Hadamard superposition state $\ket{+} = (1/\sqrt{2})(\ket{0} + \ket{1})$. }
\label{fig:decohere}
\end{figure}

Figure (\ref{fig:decohere}) shows the time evolution of the entanglement entropy  $S(\rho_S(t))$, as well as the magnitude of the off-diagonal terms in $\rho_S(t)$.
The figure is misleading in that for sufficiently large $t$ one will start to see periodic behavior again, but again this is expected since our ``Environment" Hilbert space
has only four states!
Nevertheless we start to see some of the behavior of true quantum decoherence:
\begin{itemize}
\item The off-diagonal terms of of the System density matrix $\rho_S$ written in the pointer basis  decay away on some time scale determined by $H_{\rm int}$.
For a more generic environment we expect this decay to be exponential. 
\item Because of this effect, the entanglement entropy quickly saturates at its maximum value and stays there.
\end{itemize}
The more opportunities you give your system to entangle
with multiple degrees of freedom in the environment, the more unstable you expect most quantum superposition states to be.
This is one of the reasons why you will never succeed in putting Schrodinger's cat in a superposition state - indeed one can argue that
for a macroscopic system like a cat it decoheres itself. Since these lectures are trying to avoid thought experiments, I will leave this
as an exercise for the reader.

\section{Quantum measurement}\label{sec:measure}
\begin{quote}
{\it The essentially new feature of the analysis of quantum phenomena is...the introduction of a fundamental distinction between the measuring 
apparatus and the objects under investigation}
-- Niels Bohr, 1958
\end{quote}

\begin{quote}
{\it This is clearly unsatisfactory. If quantum mechanics applies to everything then it must apply to a physicist's measurement apparatus.}
-- Steven Weinberg, {\it Lectures on Quantum Mechanics}, 2013
\end{quote}

To make a long story short, Bohr was wrong and Weinberg is correct. One of the beautiful things about having real functioning quantum computers
is that can we avoid long tedious arguments about how quantum measurements work. The quantum computers are making such measurements, so
what are they actually doing?

It has been realized for some decades now that quantum measurement is intimately related to environmentally-induced quantum decoherence.
Wojciech Zurek had the key insight that physical environments superselect certain bases, leading to pointer states \cite{Zurek:1981xq,Zurek:1982ii}.
For a two-level pointer state
this means that we have a qubit with a preferred computational basis $\ket{0}$ and $\ket{1}$. Often this preferred basis maps to energy
eigenstates. Since for a well-chosen pointer qubit superposition states are highly unstable, it performs like a classical pointer, pointing either
to 0 or to 1. 

The fact that pointer states are easy to make is the flip side of the statement that quantum computers are difficult to make. The qubits of a digitally
programmable quantum computer need to act like pointer states at the beginning and end of the quantum circuit, but in the middle of the circuit sustain highly
entangled superposition states that can be manipulated on timescales short compared to the timescale where quantum decoherence spoils the circuit.

Let us now discuss quantum measurement in the same language that we used to understand the basic features of environmentally-induced quantum decoherence. I have some pure
quantum state in the System $S$ that I would like to measure. $S$ is coupled to an Environment $E$; the new element is that $E$ contains my pointer state that
lives in some subspace $P$ of $E$. For simplicity we can take $S$ and $P$ to be single qubits, and describe what is happening in a computational basis that includes
the enviromentally-preferred bases of $S$ and $P$. 

If we make the mistake of trying to couple $P$ directly to $S$, then we will not learn anything about actual measurements, since unitary time evolution in
$S+P$ is too simple, as we saw in our two-qubit ``example" of quantum decoherence. This gross over-simplification seems to be the origin of the misleading 
but common claim that
quantum measurement cannot be understood from the point of view of entanglement and unitary evolution without assuming that the 
measurement apparatus itself is in quantum superposition. Obviously you cannot get a realistic view of quantum measurement without putting in at
least as much dynamics as you need to understand quantum decoherence, since they are so closely related. Failure to include this dynamics leads
to a lot of nonsense about ``collapse of the wavefunction", as if this were some mystical event. To put this another way, John Wheeler
liked to quote the saying ``time is Nature's way to keep everything from happening all at once"; in the same (whimsical) sense, the dynamics of quantum
decoherence is Nature's way to keep everything from being in superposition.

\subsection{Kraus operators}

Our overly-simplistic two-qubit example of $S + E$ is good enough  to introduce some basic formalism that is the key to understanding
quantum measurement in the language of quantum decoherence. Let's take the two-qubit Hamiltonian in $S + E$ to be:
\begin{eqnarray}
H_{\rm int} = -\Bigl( 
c_{11}(\sigma^z_1\otimes  \mathbbm{1}_2) 
+c_{22}( \mathbbm{1}_1 \otimes \sigma^z_2) 
+ c_{12}(\sigma^+_1\sigma^-_2 + \sigma^-_1\sigma^+_2)
 \Bigr)
\label{eq:HJC2}
\end{eqnarray}
and choose $c_{11} = c_{22} = c_{12} = 1$.
For the initial state
at $t=0$ we put the System qubit in the Hadamard state $\ket{+} = (1\sqrt{2})(\ket{0} + \ket{1})$, so
\begin{eqnarray}
\rho_S(0) = \frac{1}{2}\begin{pmatrix} 1 & 1 \\ 1 & 1 \end{pmatrix}
\end{eqnarray}
We will take the initial state of the Environment qubit $\ket{E_0}$ to be either $\ket{0}$ or $\ket{1}$.
We construct the unitary time evolution in $S+E$ as
\begin{eqnarray}
\ket{\psi(t)} = U(t)\ket{\psi(t=0)} = {\rm e}^{-iH_{\rm int}t}\;\ket{+0}{\;\rm or\;}\ket{+1}
\end{eqnarray}
and then use $\rho_{SE}(t) = U(t)\rho_{SE}(0)U^{-1}(t)$ to take the partial trace to get $\rho_S(t)$.

The new wrinkle is to ask how can we describe this unitary time evolution that produces $\rho_S(t)$ from $\rho_S(0)$ entirely in terms of
operators that act in $S$? This is the equivalent of ignoring the existence of the Environment but mocking up the resulting physics
in the single qubit System space. This is accomplished by introducing the following simple example \cite{Schlosshauer:2014pgr}
of {\it Kraus operators} acting in $S$,
constructed using the components $U_{ij}$ of the $4\times 4$ unitary time evolution operator in $S+E$:
\begin{eqnarray}
E_{11}(t) &=& 
\begin{pmatrix} 
U_{11} & U_{13} \\ U_{31} & U_{33} 
\end{pmatrix}
\\ \nonumber
E_{12}(t) &=& 
\begin{pmatrix} 
U_{12} & U_{14} \\ U_{32} & U_{34} 
\end{pmatrix}
\\ \nonumber
E_{21}(t) &=& 
\begin{pmatrix} 
U_{21} & U_{23} \\ U_{41} & U_{43} 
\end{pmatrix}
\\ \nonumber
E_{22}(t) &=& 
\begin{pmatrix} 
U_{22} & U_{24} \\ U_{42} & U_{44} 
\end{pmatrix}
\label{eq:Kraus}
\end{eqnarray}

You can then verify the following expressions for  $\rho_S(t)$:
\begin{eqnarray}
 \ket{E_0} = \ket{0} \; : \quad \rho_S(t) &=& E_{11}.\rho_S(0).E_{11}^\dagger +  E_{21}.\rho_S(0).E_{21}^\dagger  \\\nonumber
 \ket{E_0} = \ket{1} \; : \quad \rho_S(t) &=& E_{12}.\rho_S(0).E_{12}^\dagger +  E_{22}.\rho_S(0).E_{22}^\dagger 
\end{eqnarray}

The $2\times 2$ Kraus operators are not unitary; in fact they are not even Hermitian. So from this truncated point of view that only allows
operations in $S$, what appears to be going on? To see this, let's define the following bilinear combinations of the Kraus operators:
\begin{eqnarray}
P_{11} \equiv E_{11}.E_{11}^\dagger \; ; \quad P_{21} \equiv E_{21}.E_{21}^\dagger \\\nonumber
P_{12} \equiv E_{12}.E_{12}^\dagger \; ; \quad P_{22} \equiv E_{22}.E_{22}^\dagger 
\end{eqnarray}
Now let's look at some numerical values for the case  $\ket{E_0} = \ket{0}$ and at time $t=1$:
\begin{eqnarray}
P_{11} = 
\begin{pmatrix} 1 & 0 \\ 0 & 0.29 \end{pmatrix} \; ;\quad
P_{12} = 
\begin{pmatrix} 0 & 0 \\ 0 & 0.71 \end{pmatrix} \; ; \quad
P_{11} + P_{12} =
\begin{pmatrix} 1 & 0 \\ 0 & 1 \end{pmatrix} \\
P_{11}^2 = 
\begin{pmatrix} 1 & 0 \\ 0 & 0.09 \end{pmatrix} \; ;\quad
P_{12}^2 = 
\begin{pmatrix} 0 & 0 \\ 0 & 0.50 \end{pmatrix} \; ; \quad
P_{11}.P_{12} = 
\begin{pmatrix} 0 & 0 \\ 0 & 0.21 \end{pmatrix} \\
P_{21} = 
\begin{pmatrix} 0.71 & 0 \\ 0 & 0 \end{pmatrix} \; ;\quad
P_{22} = 
\begin{pmatrix} 0.29 & 0 \\ 0 & 1 \end{pmatrix} \; ; \quad
P_{21} + P_{22} =
\begin{pmatrix} 1 & 0 \\ 0 & 1 \end{pmatrix} \\
P_{21}^2 = 
\begin{pmatrix} 0.5 & 0 \\ 0 & 0 \end{pmatrix} \; ;\quad
P_{22}^2 = 
\begin{pmatrix} 0.09 & 0 \\ 0 & 1 \end{pmatrix} \; ; \quad
P_{21}.P_{22} = 
\begin{pmatrix} 0.21 & 0 \\ 0 & 0 \end{pmatrix}
\end{eqnarray} 

Again this oversimplified example displays simple periodic behavior rather than quantum decoherence, but the dynamics
does produce nonzero entanglement entropy, and in fact for $t=1$ we find $S(\rho_S(t))= 0.3$.
Looking at the operators $P_{ij}$, it is pretty obvious what they are doing: each set $P_{11}$,$P_{12}$ and
$P_{21}$,$P_{22}$ is acting {\it approximately} like the $2\times 2$ {\it projection operators} $P^0$,$P^1$ onto the
System computational basis.

This is the basic trick whereby the environmental dynamics responsible for quantum coherence resembles the projection operators
that we introduce out of nowhere when invoking the so-called Measurement Postulate of quantum mechanics.
{\it What we call quantum measurement is a special case of this more general dynamics}. To see it work in conjunction with the
phenomena of pointer states, you need to include a realistic number of degrees of freedom, as I have already emphasized.

\subsection{The SWAP operator and ``collapse" of the wavefunction}

As a last example of how we often get confused about quantum measurement, let's consider the simple SWAP quantum circuit shown in
Figure~\ref{fig:Exp3}, that we discussed already in Section \ref{sec:two}.
In this circuit we prepared the qubit 0 in the superposition state $\ket{+} = (1/\sqrt{2})(\ket{0} + \ket{1})$. We now want to measure this
state by the following two-step process:
\begin{itemize}
\item Transfer the quantum state information from qubit 0 to qubit 1 using the combination of three {\bf CNOT} operators as shown in the figure.
\item Measure qubit 1 in the computational basis to obtain the
state information originally carried by qubit 0. 
\end{itemize}

%

It is important to keep in mind that it is the combination of these two steps that is the measurement.
This is a perfectly normal way of measuring qubit 0, and when repeated many times should give 0 half of the time and 1 half of the time,
consistent with the fact that we prepared qubit 0 in the state $\ket{+}$. Now ask the question: suppose that after this measurement is complete,
we re-measure qubit 0 in the computational basis; this corresponds precisely to the upper dial in Figure~\ref{fig:Exp3}.
What results will we get? In the usual sloppy way that we teach quantum mechanics, we say
that the second measurement of qubit 0 in the computational basis should give the same result as the first measurement in the computational basis,
because after the first measurement the ``wavefunction of qubit 0 has collapsed". To see that this is false, recall the results that we got
from running the notebook:

\begin{verbatim}
Cirq circuit:

0: ───H─────@───X───@─────────────M('q0')───
            │   │   │
1: ───X^t───X───@───X───M('q1')─────────────

Results for t = 0:

q0=00000000000000000000000000000000000000000000000000
q1=01100011111010111111111110101101010100011010111001

Results for t = 1:

q0=11111111111111111111111111111111111111111111111111
q1=10100111011100100101011100101100100001001000101100

Results for t = 0.5:

q0=10100101111000101011000101100101100010100011110111
q1=10111101111001000110010010000011011001000111100111
\end{verbatim}

As you see, for three different choices of a parameter that has to do with the physical state of the measuring apparatus before the first measurement occurs,
the results of the two independent measurements of the state of qubit 0 give completely uncorrelated results. This is because the SWAP operation,
in the process of swapping out the quantum information from qubit 0 in order to measure it, also swaps in a particular (but arbitrary) state. Thus the second
measurement sees this other state, not the "collapsed" version of the original state.


\section{Bell inequalities}\label{sec:Bellinequality}
In 1964, a CERN particle theorist named John Stewart Bell wrote a remarkable paper titled ``On the Einstein-Podolsky-Rosen Paradox" \cite{Bell:1964kc}.
In this paper and
some follow-on work, he derived the famous Bell inequalities. By obtaining measurements of expectation values in the laboratory that violate a Bell
inequality, you can prove that your system is indeed a quantum system with quantum entanglement, not a classical system with clever correlations built into
a ``hidden" classical configuration space. Here is a simple example of how this works.

Let's generalize the simple EPR pair to a one-parameter family of 2-qubit entangled states:
\begin{eqnarray}
\ket{\psi} = {\rm cos}\,\alpha \ket{00} + {\rm sin}\,\alpha\ket{11}
\label{eq:genent}
\end{eqnarray}
Obviously for $\alpha = 0$ or $\pi/2$ the two qubits are not entangled, but for any other value $0 < \alpha < \pi/2$ there is some amount of entanglement.
We can measure this by computing the entanglement entropy. Starting with the density matrix:
\begin{eqnarray}
\rho_{12} = 
\begin{pmatrix}
{\rm cos}^2\alpha & 0 & 0 & {\rm cos}\,\alpha\, {\rm sin}\,\alpha \\
0 & 0 & 0 & 0 \\
0 & 0 & 0 & 0 \\
{\rm cos}\,\alpha \,{\rm sin}\,\alpha & 0 & 0 & {\rm sin}^2\alpha 
\end{pmatrix}
\end{eqnarray}
If we take the partial trace with respect to either the first or second qubit, we get
\begin{eqnarray}
\rho_1 = \rho_2 = 
\begin{pmatrix}
{\rm cos}^2\alpha & 0 \\
 0 & {\rm sin}^2\alpha
\end{pmatrix}
\end{eqnarray}
which gives the entanglement entropy
\begin{eqnarray}
S_1 = S_2 = -\left( {\rm cos}^2 \alpha \,{\rm log}\,( {\rm cos}^2\alpha )+  {\rm sin}^2\alpha \,{\rm log}\,( {\rm sin}^2\alpha ) \right)
\end{eqnarray}
As expected, $S_1$ vanishes for $\alpha = 0$ or $\pi/2$, and has a maximum value of 1 when $\alpha = \pi/4$, corresponding to an EPR pair.

Now suppose that Carmen prepares a large number of photon pairs in the state $\ket{\psi}$ with some fixed value of $\alpha$.
She sends one photon from each pair to Alice, and the other to Bob. For each photon that Alice receives, she  chooses at random
whether to measure it the basis where $\sigma_1^z$ is diagonal (the usual computational basis) or in the basis where $\sigma_1^x$
is diagonal (the Hadamard basis). Since both $\sigma_1^z$ and $\sigma_1^x$ have eigenvalues $\pm 1$, we can denote the results
of these measurements collectively by $Q$ and $R$, where each $q\in Q = \pm 1$, and each $r\in R = \pm 1$.

Bob does the same thing with his photons, but chooses two different bases, specified as the bases in which the following combinations of
Pauli matrices are diagonal:
\begin{eqnarray}
{\rm cos}\,\beta\,\sigma_2^z &+& {\rm sin}\,\beta\, \sigma_2^x \\
{\rm cos}\,\beta'\sigma_2^z &+& {\rm sin}\,\beta' \sigma_2^x 
\end{eqnarray}
 We can denote the results
of these measurements collectively by $S$ and $T$, where each $s\in S = \pm 1$, and each $t\in T = \pm 1$.

If we know the state $\ket{\psi}$, we can use quantum mechanics to predict the following joint expectation values:
\begin{eqnarray}
\langle QS \rangle &=& {\rm cos}\,\beta \bra{\psi}\sigma_1^z \sigma_2^z \ket{\psi} + {\rm sin}\,\beta \bra{\psi}\sigma_1^z \sigma_2^x \ket{\psi} \\
&=& {\rm cos}\,\beta \left( {\rm cos}^2\alpha + {\rm sin}^2\alpha \right) \\
&=& {\rm cos}\,\beta\\
\langle QT \rangle &=& {\rm cos}\,\beta' \\
\langle RS \rangle &=&  {\rm cos}\,\beta \bra{\psi}\sigma_1^x \sigma_2^z \ket{\psi} + {\rm sin}\,\beta \bra{\psi}\sigma_1^x \sigma_2^x \ket{\psi} \\
&=& 2{\rm cos}\,\alpha\, {\rm sin}\,\alpha \,{\rm sin}\,\beta \\
&=& {\rm sin}\,2\alpha \,{\rm sin}\,\beta \\
\langle RT \rangle &=&  {\rm sin}\,2\alpha \,{\rm sin}\,\beta'
\end{eqnarray}

Now consider the following linear combination of expectation values (please note the minus sign):
\begin{eqnarray}
E_{\rm Bell} \equiv \langle QS \rangle + \langle RS \rangle + \langle RT \rangle - \langle QT \rangle = {\rm cos}\,\beta - {\rm cos}\,\beta' + {\rm sin}\,2\alpha \,\left( {\rm sin}\,\beta + {\rm sin}\,\beta' \right)
\end{eqnarray}
The choice of $\beta'$ that maximizes $E_{\rm Bell}$ is obviously ${\rm cos}\,\beta' = -{\rm cos}\,\beta$, ${\rm sin}\,\beta' = {\rm sin}\,\beta > 0$.
Then we have:
\begin{eqnarray}
E_{\rm Bell} = 2{\rm cos}\,\beta + 2{\rm sin}\,2\alpha \,{\rm sin}\,\beta
\end{eqnarray}
The choice of $\beta$ that then maximizes $E_{\rm Bell}$ is given by (check this yourself):
\begin{eqnarray}
{\rm cos}\,\beta = \frac{1}{\sqrt{1+{\rm sin}^2 2\alpha}}\;,\quad
{\rm sin}\,\beta = \frac{{\rm sin}\,2\alpha}{\sqrt{1+{\rm sin}^2 2\alpha}}
\label{eq:optbasis}
\end{eqnarray}
which gives:
\begin{eqnarray}
E_{\rm Bell} = 2\sqrt{1+{\rm sin}^2 2\alpha}
\end{eqnarray}
Notice that, for any value of $\alpha$, $0 < \alpha < \pi/2$, the quantum prediction is $E_{\rm Bell} > 2$.

Now let's compare this quantum prediction for $E_{\rm Bell}$ with the prediction of the diehard classical physicist, who tries to describe the system in terms of a
``hidden variable" theory with ``local realism". There is a whole literature on the assumptions and caveats of this sort of thinking that we can happily ignore.
I only need to assume that, before the measurements are performed, we have system variables $q$, $r$, $s$, $t$, all taking values $\pm 1$,  in a classical configuration space
with a prior probability distribution given by some function $\cal{P}$$(q,r,s,t)$. In this classical language the measurements $Q$, $R$, $S$, $T$ are
just sampling from this distribution. We assume that Alice and Bob's measurements follow the classical protocol that we described in subsection (\ref{ss:cmt}),
and we assume that Bob's measurement results are not affected by Alice's measurment results other than through the correlations inherent in $\cal{P}$$(q,r,s,t)$.

Lacking an actual hidden variable theory, our classical physicist has no way of knowing the prior probability distribution $\cal{P}$$(q,r,s,t)$, but can easily
derive an inequality for $E_{\rm Bell}$ that must be obeyed for any  $\cal{P}$$(q,r,s,t)$. Notice that
\begin{eqnarray}
qs + rs + rt -qt = (r+q)s + (r-q)t
\end{eqnarray}
Since all the system variables take values $\pm 1$, we see that in every case either $(r+q)s = 0$ or $(r-q)t = 0$, and that in all cases
$qs + rs + rt -qt = \pm2$. So we can write the inequality:
\begin{eqnarray}
E_{\rm Bell} &=& \sum_{q,r,s,t = \pm1} \, {\cal P}(q,r,s,t) \, \left( qs + rs + rt + qt \right) \\
&\leq& 2  \sum_{q,r,s,t = \pm1} \,{\cal  P}(q,r,s,t) = 2
\end{eqnarray}
Thus this classical reasoning produces an example of a Bell inequality:
\begin{eqnarray}
E_{\rm Bell} \leq 2
\end{eqnarray}
Actually this particular version was dervived by Clauser, Horne, Shimony, and Holt, and is known as the CHSH inequality \cite{Clauser:1969ny}.

\begin{figure}
  \centering
    \includegraphics[width=0.75\textwidth]{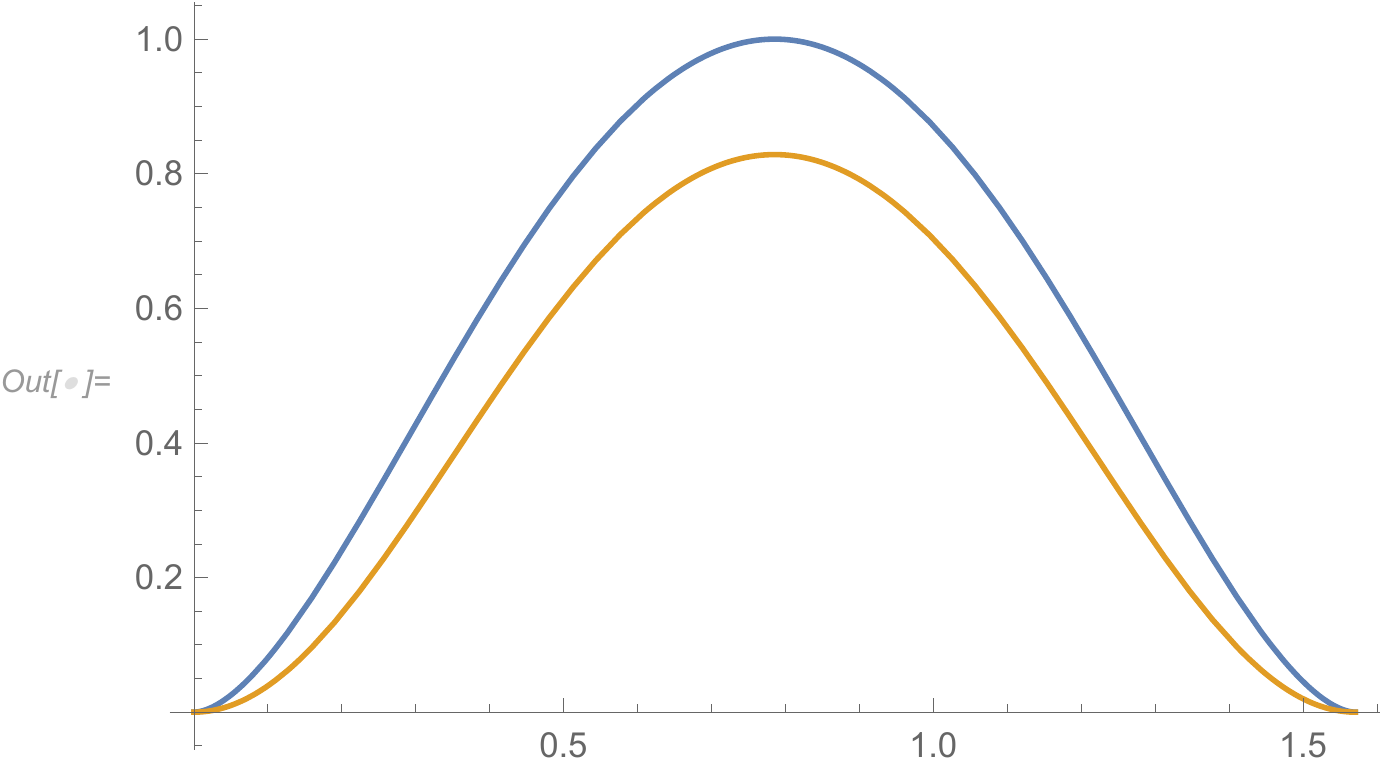}
 \caption{Entanglement entropy $S_1$ (blue) and violation of the Bell inequality $EB-2$ (orange), plotted as a function of $\alpha$.}
\label{fig:BEvsS}
\end{figure}

Now we see that, for suitably chosen measurement protocols, the quantum state $\ket{\psi}$ can produce experimental results
that violate the Bell inequality, for some amount of entanglement in the prepared state. For example, Bob could choose
${\rm cos}\,\beta = -{\rm cos}\,\beta' = {\rm sin}\,\beta = {\rm sin}\,\beta' = \sqrt{2}/2$, in which case Alice and Bob together will
measure:
\begin{eqnarray}
E_{\rm Bell} = \sqrt{2} \left( 1 + {\rm sin}\,2\alpha \right)
\end{eqnarray}
which will violate the Bell inequality for any value of $\alpha > 12$ degrees.
If Alice, Bob, and Carmen conspire together to choose optimal bases as given
by Eq. (\ref{eq:optbasis}),  then the Bell inequality will be violated by an amount given by the function
\begin{eqnarray}
E_{\rm Bell} - 2 = 2\left( \sqrt{1 + {\rm sin}^2 2\alpha} - 1 \right)
\end{eqnarray}
In Figure (\ref{fig:BEvsS}) we see this quantity, together with $S_1$, plotted as a function of $\alpha$. 
They have very similar functional forms, since in fact {\it the violation of the Bell inequality is an experimental
measurement of the quantum entanglement} in the two-photon system. Such laboratory measurements have in fact been
made many times, forcing us to accept the reality of the quantum description.

\section{Area law for entanglement entropy in quantum field theory}\label{sec:six}
In 1993 Mark Srednicki wrote a short paper titled {\it Entropy and Area} \cite{Srednicki:1993im}.
It is one of those papers that manages to be both simple
and profound. It is so simple that I can (and will) reproduce its entire content for you now; it is so profound that its true message has only partly
been appreciated by the particle physics community some 27 years later. The only physical systems considered in this paper are quantum harmonic
oscillators; how could we possibly learn anything new, let alone profound, by poking around in such familiar territory? 

Here is what we will demonstrate:
\begin{itemize}
\item A direct mapping between entanglement entropy and Shannon entropy via a quantum construction of a {\it thermofield double state}.
\item An area law for entanglement entropy in quantum field theory, that has nothing to do with black holes - or does it?
\end{itemize}

\subsection{Shannon entropy for a thermal ensemble of quantum oscillators}
Consider $N$ uncoupled harmonic oscillators; they can be described by a free Hamiltonian:
\begin{eqnarray}
H = \sum_{i=1}^N \,H_i = \frac{1}{2} \sum_{i=1}^{N} \left( \pi_i^2 + \omega_i^2\phi_i^2 \right)
\end{eqnarray}
where each oscillator is quantized and can thus be represented by creation/annihilation operators $a_i^\dagger$,$a_i$ via:
\begin{eqnarray}
\phi_i &=& \frac{1}{\sqrt{2\omega_i}}\left( a_i^\dagger + a_i \right) \\\nonumber
\pi_i &=& \sqrt{\frac{\omega_i}{2}}\, i \,\left( a_i^\dagger - a_i \right) 
\end{eqnarray}
We have the usual energy eigenstates labelled by the nonnegative eigenvalues $n$ of the number operator $a_i^\dagger a_i$:
\begin{eqnarray}
\ket{n}_i = \frac{1}{\sqrt{n!}}(a_i^\dagger)^n \ket{0}_i
\end{eqnarray}

Suppose that for each oscillator we prepare a Boltzmann thermal ensemble characterized by some temperature $kT_i = 1/\beta_i$.
Then we can define a density matrix $\rho_i$ for this mixed state of each oscillator; this matrix is diagonal in the energy eigenbasis:
\begin{eqnarray}
\rho^i_{nm} = \delta_{nm}\frac{1}{Z_i} {\rm e}^{-\beta_i\omega_i(n+1/2)}
\end{eqnarray}
Where $Z_i$ is the partition function of the oscillator and the factor of $(1/Z_i)$ ensures that $\bm{tr}(\rho_i) = 1$:
\begin{eqnarray}
Z_i = \sum_{n=0}^{\infty} \,  {\rm e}^{-\beta_i\omega_i(n+1/2)} = \frac{  {\rm e}^{-\beta_i\omega_i /2}}{1 -  {\rm e}^{-\beta_i\omega_i }}
= \frac{1}{2\,{\rm sinh}\left( \frac{\beta_i\omega_i}{2} \right)}
\label{eq:partf}
\end{eqnarray}

The Shannon entropy of each oscillator can be computed as:
\begin{eqnarray}
S(\rho^i) &=& -\bm{tr}(\rho^i {\rm log}\,\rho^i )\\\nonumber
&=& -\sum_{n=0}^{\infty} \left( 1-  {\rm e}^{-\beta_i\omega_i } \right) \,  {\rm e}^{-\beta_i\omega_i n } 
{\rm log}\left[  \left( 1-  {\rm e}^{-\beta_i\omega_i } \right) \,  {\rm e}^{-\beta_i\omega_i n } \right] \\\nonumber
&=& -{\rm log} \left( 1-  {\rm e}^{-\beta_i\omega_i } \right) + \frac{\beta_i\omega_i  {\rm e}^{-\beta_i\omega_i }}
{ \left( 1-  {\rm e}^{-\beta_i\omega_i } \right)}
\end{eqnarray}
This result can be rewritten in an equivalent form that will be useful later:
\begin{eqnarray}
S(\rho^i) = (c_i + 1/2){\rm log}(c_i+ 1/2) -  (c_i - 1/2){\rm log}(c_i- 1/2) 
\label{eq:Swithc}
\end{eqnarray}
where we have used the notation
\begin{eqnarray}
c_i &=& \frac{1}{2} \frac{ \left( 1+  {\rm e}^{-\beta_i\omega_i } \right)}{ \left( 1-  {\rm e}^{-\beta_i\omega_i } \right)} \\\nonumber
&=& \frac{1}{2}\, {\rm coth}\left(\frac{\beta_i\omega_i}{2}\right)
\end{eqnarray}
We can relate the $c_i$ to the two-point correlators of each oscillator. Using expectation values to denote ensemble averages, we have
for each decoupled oscillator:
\begin{eqnarray}
\bm{tr}\left( \rho^i \, \phi_i \phi_i \right) &=& \langle \phi_i \, \phi_i \rangle \\\nonumber
  &=& \frac{1}{2\omega_i} \langle a_i^\dagger a_i + a_i a_i^\dagger \rangle = \frac{1}{2\omega_i} \langle 2a_i^\dagger a_i + 1 \rangle \\\nonumber
&=& \frac{1}{\omega_iZ_i} \sum_{n=0}^{\infty}  {\rm e}^{-\beta_i\omega_i(n+1/2) } \, (n+1/2) \\\nonumber
&=& \frac{1}{2\omega_i}\, {\rm coth}\left( \frac{\beta_i\omega_i}{2} \right) \\
\bm{tr}\left( \rho^i \, \pi_i \pi_i \right) &=& \langle \pi_i \, \pi_i \rangle \\\nonumber
&=& \frac{\omega_i}{2}\, {\rm coth}\left( \frac{\beta_i\omega_i}{2} \right)
\end{eqnarray}
So we can connect the entropy formula to these correlators via the expression
\begin{eqnarray}
c_i = \sqrt{ \langle \phi_i \, \phi_i \rangle \langle \pi_i \, \pi_i \rangle}
\label{eq:cdef}
\end{eqnarray}

\subsection{A thermofield double from coupled oscillators}
Now let's consider oscillators with a Gaussian coupling. Instead of introducing a classical thermal ensemble by hand,
we are going to look at what happens in the quantum system when we trace out over some of the quantum oscillators.
For general $N$ we write the Hamiltonian as
\begin{eqnarray}
H = \frac{1}{2} \sum_{i=1}^{N} \pi_i^2 + \frac{1}{2}  \sum_{i,j=1}^{N} \phi_i K_{ij} \phi_j
\label{eq:HK}
\end{eqnarray}
where the coupling matrix $K_{ij}$ is real and symmetric. Of course since these are Gaussian couplings we
can always find a new basis with decoupled oscillators; using
\begin{eqnarray}
K_{ij} = O^T_{ik} \omega_k O_{kj}
\end{eqnarray}
we can write an equivalent Hamiltonian
\begin{eqnarray}
H =   \frac{1}{2} \sum_{i=1}^{N} \left( \tilde{\pi}_i^2 + \omega_i^2\tilde{\phi}_i^2 \right) \; ; \quad
\tilde{\pi}_i = O_{ij}\pi_j \; ;\quad \tilde{\phi}_i = O_{ij}\phi_j 
\end{eqnarray}

Let's consider first the simplest case $N=2$, and label the oscillators $\phi_L$, $\phi_R$. We introduce an explicit notation for the $2\times 2$ coupling matrix:
\begin{eqnarray}
K_{ij} = \omega^2
\begin{pmatrix}
1 + 2\,{\rm tan}^2\theta & 2\,{\rm tan}\,\theta / {\rm cos}\,\theta \\
 2\,{\rm tan}\,\theta / {\rm cos}\,\theta & 1 + 2\,{\rm tan}^2\theta 
\end{pmatrix}
\end{eqnarray}
where the parameter $\theta$ determines the strength of the coupling between the two oscillators.
Diagonalizing $K_{ij}$ produces two eigenvalues:
\begin{eqnarray}
\omega^2_{\pm}  = \omega^2 \frac{(1\pm {\rm sin}\,\theta )^2}{{\rm cos}^2\theta}
\; ; \quad
\tilde{\phi}_\pm = \frac{1}{\sqrt{2}}(\phi_l \pm \phi_r ) \; ; \quad
\tilde{\pi}_\pm = \frac{1}{\sqrt{2}}(\pi_l \pm \pi_r )
\end{eqnarray}
Let $a_L^\dagger$,$a_L$, $a_R^\dagger$,$a_R$ denote the creation/annihilation operators of the decoupled system with $\theta = 0$
and energy eigenstates $\ket{n}_L\ket{n}_R$. You can construct the ground state of the coupled system by starting with the ansatz:
\begin{eqnarray}
\ket{0} = \frac{1}{\sqrt{\cal N}} {\rm e}^{A a_L^\dagger a_R^\dagger} \ket{0}_L\ket{0}_R \; ; \quad {\cal N} = \frac{1}{1-A^2}
\end{eqnarray}
and computing $A$ by requiring that $\ket{0}$ is annihilated by the annihilation operators of the rotated basis:
\begin{eqnarray}
a_\pm &=& \frac{1}{\sqrt{2}} \left( \sqrt{\omega} \tilde{\phi}_{\pm} + \frac{i}{\sqrt{\omega}}\tilde{\pi}_{\pm} \right) \\\nonumber
a_\pm \ket{0} &=& 0
\end{eqnarray}
You should obtain the following result:
\begin{eqnarray}
A = -{\rm tan}\,\frac{\theta}{2}
\end{eqnarray}

Now we can rewrite the ground state in the following suggestive form by making the substitution $|A| \to $ exp$(-\beta\omega/2)$:
\begin{eqnarray}\label{eq:TFD2}
\ket{0} &=&  \frac{1}{\sqrt{\cal N}} {\rm e}^{A a_L^\dagger a_R^\dagger} \ket{0}_L\ket{0}_R \\\nonumber
&=&  \frac{1}{\sqrt{\cal N}} \sum_{n=0}^{\infty} \frac{A^n}{n!} (a_L^\dagger)^n  (a_R^\dagger)^n  \ket{0}_L\ket{0}_R \\\nonumber
&=&  \frac{1}{\sqrt{\cal N}} \sum_{n=0}^{\infty} A^n \ket{n}_L\ket{n}_R \\\nonumber
&=&   \frac{1}{\sqrt{\cal N}}\frac{1}{|A|} \sum_{n=0}^{\infty} {\rm e}^{-\beta E_{nn}/2} \ket{n}_L\ket{n}'_R \\\nonumber
&=&   \frac{1}{\sqrt{Z}}\sum_{n=0}^{\infty} {\rm e}^{-\beta E_{nn}/2} \ket{n}_L\ket{n}'_R
\end{eqnarray}
where $E_{nn} = (2n+1)\omega$ are the energies of the uncoupled $\theta = 0$ basis states $ \ket{n}_L\ket{n}'_R$,
and we have written the right-side eigenstates with a different phase convention: $\ket{n}'_R \equiv (-1)^n\ket{n}_R$.

The ground state is a pure state, but we have written it in a form were it is starting to resemble a thermal ensemble.
For this reason the final expression in Eq. (\ref{eq:TFD2}) is called a {\it thermofield double state} or TFD \cite{Maldacena:2001kr,Shenker:2013pqa}.
Of course the TFD that we have derived is not a thermal state at all, but rather a pure state with entanglement
between the left and right oscillators. The density matrix of this pure state is just
\begin{eqnarray}
\rho_0 =  \frac{1}{\cal N} \sum_{n,m=0}^{\infty} A^{n+m} \ket{n}_L\ket{n}_R \bra{n}_L\bra{n}_R 
\end{eqnarray}

Now let's take a partial trace of $\rho_0$ over the right-side degrees of freedom:
\begin{eqnarray}
 \rho_L = \bm{tr_R}\ket{0}\bra{0} &=&  \frac{1}{\cal N} \sum_{n=0}^{\infty} A^{2n} \ket{n}_L \bra{n}_L \\\nonumber
&=& \left( 1 - {\rm e}^{-\beta\omega} \right) \sum_{n=0}^{\infty} \left( {\rm e}^{-\beta\omega n} \right) \ket{n}_L\bra{n}_L \\\nonumber
&=& \frac{1}{Z} \sum_{n=0}^{\infty} {\rm e}^{-\beta E_n}  \ket{n}_L\bra{n}_L
\end{eqnarray}
where $E_n = (n+1/2)\omega$ and $Z= 1/2{\rm sinh}( \frac{\beta\omega}{2})$, as in Eq. (\ref{eq:partf}).

Of course $\rho_L$ describes a mixed state, but it is not just any mixed state: it is precisely what one would obtain from a Boltzmann
ensemble with temperature given by
\begin{eqnarray}
T = -\frac{\omega}{2\,{\rm log}\left[{\rm tan}\frac{\theta}{2} \right]}
\end{eqnarray}

%
\begin{figure}[tb]
\centering
\includegraphics[width=0.65\textwidth]{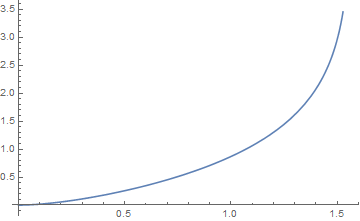}
\caption{Entanglement entropy $S(\rho_L)$ of oscillator $\phi_L$ with respect to tracing out the oscillator $\phi_R$ that couples to it,
as a function of the coupling parameter $\theta$. The logarithmic divergence as $\theta \to \pi/2$ corresponds to the ``temperature"
$T \to \infty$.
\label{fig:osc}}
\end{figure}
%

At this point we can repeat the analysis of the previous subsection to compute the entanglement entropy. Using
\begin{eqnarray}
\langle \tilde{\phi}_{\pm} \, \tilde{\phi}_{\pm} \rangle &=& \frac{1}{2\omega_{\pm}} \label{eq:phitc} \\
\langle \tilde{\pi}_{\pm} \, \tilde{\pi}_{\pm} \rangle &=& \frac{\omega_{\pm}}{2} \label{eq:pitc} \\
 \langle \phi_L \, \phi_L \rangle &=& \frac{1}{2}\left( \frac{1}{2\omega_+} + \frac{1}{2\omega_-} \right) \label{eq:phic}
 \\
 \langle \pi_L \, \pi_L \rangle &=&  \frac{1}{2}\left( \frac{\omega_+}{2} + \frac{\omega_-}{2} \right)  \label{eq:pic}
\end{eqnarray}
then using the fact that $\omega_{\pm} = \omega (1 \pm {\rm sin}\,\theta )/{\rm cos}\,\theta$, we find
\begin{eqnarray}
c &=&  \sqrt{ \langle \phi_L \, \phi_L \rangle \langle \pi_L \, \pi_L \rangle}
= \left[ \frac{1}{16} \left( \frac{1}{\omega_+} + \frac{1}{\omega_-} \right)\left( \omega_+ + \omega_- \right) \right]^{1/2} \\\nonumber
&=& \frac{1}{2\,{\rm cos}\,\theta}
\end{eqnarray}
Now we use Eq. (\ref{eq:Swithc}) to compute the entropy:
\begin{eqnarray}
S(\rho_L) &=& \frac{1}{2}( {\rm sec}\,\theta + 1){\rm log}\left[\frac{1}{2}( {\rm sec}\,\theta + 1)\right]
- \frac{1}{2}( {\rm sec}\,\theta - 1){\rm log}\left[\frac{1}{2}( {\rm sec}\,\theta - 1)\right] \\\nonumber
&=&  -{\rm log}\left[ 1 - {\rm tan}^2\frac{\theta}{2} \right] - \frac{ {\rm tan}^2\frac{\theta}{2} }{ 1 - {\rm tan}^2\frac{\theta}{2}} 
{\rm log}\left[ {\rm tan}^2\frac{\theta}{2} \right] \\\nonumber
&=& -{\rm log} \left( 1-  {\rm e}^{-\beta\omega } \right) + \frac{\beta\omega  {\rm e}^{-\beta\omega }}
{ \left( 1-  {\rm e}^{-\beta\omega } \right)}
\end{eqnarray}
where to get to the final line we used $ {\rm tan}^2\frac{\theta}{2} = {\rm exp}(-\beta\omega )$, and to get to the second line
we used the trigonometric identities:
\begin{eqnarray}
\frac{1}{2}( {\rm sec}\,\theta + 1) = \frac{1}{1- {\rm tan}^2\frac{\theta}{2}}\; ; \quad
\frac{1}{2}( {\rm sec}\,\theta - 1) = \frac{ {\rm tan}^2\frac{\theta}{2}}{1- {\rm tan}^2\frac{\theta}{2}}
\end{eqnarray}

Keep in mind that what we have computed here is the entanglement entropy of a quantum oscillator entangled with another oscillator.
For weak coupling -- small values of $\theta$ or low ``temperature" $T$ -- the dynamics is similar to our discussion of quantum decoherence, and even better than our
3-qubit example since here we have a Hilbert space with an infinite number of states.
For larger $\theta$ the behavior of the entropy is dominated by the logarithmic divergence at $\theta \to  \pi/2$,
which corresponds to the ``temperature" $T \to \infty$.
In Figure \ref{fig:osc} you can compare the exact formula for the entropy to the following approximate formula that captures the log divergence:
\begin{eqnarray}
S(\rho_L) \simeq -{\rm log}\,\epsilon + 1 - \frac{1}{2}\epsilon \; ; \quad \epsilon = 1- {\rm tan}^2\frac{\theta}{2} \simeq \frac{\omega}{T}
\end{eqnarray}

\subsection{Area law for entanglement entropy of scalar fields}

Having gone to a lot of trouble to set up a nice formalism to treat the case of $N=2$ oscillators, it is now easy to consider the general case.
In fact let's go directly to the case of a free massless scalar field theory in $3+1$ dimensions:
\begin{eqnarray}
H = \frac{1}{2} \int d^3x \left( \pi^2(x) + (\nabla\phi(x))^2 \right)
\end{eqnarray}
We will need to introduce both an infrared and an ultraviolet cutoff in order to reduce the problem to an arbitrarily large but discrete set of 
oscillators. As a first step introduce the usual partial wave expansion:
\begin{eqnarray}
\phi_{lm}(r) &=& r\, \int d\Omega\, \phi(x) \, Y_{lm}(\theta ,\varphi ) \\\nonumber
\pi_{lm}(r) &=& r\, \int d\Omega\, \pi(x) \, Y_{lm}(\theta ,\varphi )
\end{eqnarray}
where the $Y_{lm}(\theta ,\varphi )$ are spherical harmonics. 
The Hamiltonian becomes:
\begin{eqnarray}
H &=& \sum_{l=0}^{\infty}\sum_{m=-l}^{l} \,H_{lm} \\\nonumber
H_{lm} &=& \frac{1}{2}\int_0^{\infty} dr\, \left( \pi_{lm}^2(r) + r^2 \left[\frac{\partial}{ \partial r} \left( \frac{\phi_{lm}(r)}{r} \right) \right]^2 
+ \frac{l(l+1)}{r^2} \phi_{lm}^2(r) \right)
\end{eqnarray}
Eventually we will truncate the sum over angular momentum eigenvalues $l$, which
is a kind of ultraviolet cutoff.

We still need to discretize the radial dependence of $\phi_{lm}(r)$ and $\pi_{lm}(r)$, in order to have a finite number of oscillators. 
Let's restrict $r$ to only take values $\frac{3}{2}a$, $\frac{5}{2}a,\ldots (N-1/2)a$, $(N + 1/2)a$, where $a$ is a radial lattice spacing (an ultraviolet cutoff)
and $N$ is a large integer (an infrared cutoff). Thus we are now confined to a sphere of radius $R = (N+1/2)a$. Measuring everything in units of $a$, we now have
a system of coupled oscillators with a Hamiltonian in the form of Eq. (\ref{eq:HK}):
\begin{eqnarray}
H &=& \frac{1}{2} \sum_{i=1}^{N} \pi_i^2 + \frac{1}{2}  \sum_{i,j=1}^{N} \phi_i K_{ij} \phi_j \\\nonumber
K_{ij} &=& \frac{ \delta_{ij}}{j^2} \left[  (j+1/2)^2 + (j-1/2)^2  + l(l+1) \right]
-(\delta_{i,j+1} + \delta_{i+1,j}) \left(  \frac{(j+1/2)^2}{j(j+1)} \right) 
\end{eqnarray}
where I have been a little sloppy about the endpoints and for the moment am suppressing the $l,m$ dependence.

Now let's divide our system of oscillators into two subsystems $A$ and $B$: we define $A$ to consist of the $\phi_j$ with $j > j_{max}$, and
the system $B$ to consist of the $\phi_j$ with $j \leq j_{max}$. Thus system $B$ consists of the oscillators inside the radius $R_B = (j_{max} + 1/2)a$,
and system $A$ is the oscillators in the spherical shell between $R_B$ and $R$. Now we imagine do the partial trace of the density matrix $\rho_{A+B}$
over the subsystem $B$, giving $\rho_A = \bm{tr_B}(\rho_{A+B})$, and computing the resulting entanglement entropy $S(\rho_A)$.

This is easy to do numerically. We start by diagonalizing $K_{ij}$ 
to get its $N$ real positive eigenvalues $\omega_j^2$. We can then write the two-point correlators in the basis where
$K_{ij}$ is diagonal, i.e. the basis where the oscillators decouple, in analogy to Eqs (\ref{eq:phitc}) and  (\ref{eq:pitc}):
\begin{eqnarray}
\langle \tilde{\phi}_j \, \tilde{\phi}_j \rangle = \frac{1}{2\omega_j} \; ; \quad
\langle \tilde{\pi}_j \, \tilde{\pi}_j \rangle = \frac{\omega_j}{2} \label{eq:genpitc} 
\end{eqnarray}
Then the matrix analogs of Eqs. (\ref{eq:phic}) and  (\ref{eq:pic}) are
\begin{eqnarray}
 \langle \phi_i \, \phi_j \rangle &=& \frac{1}{2}\left( K^{-1/2} \right)_{ij} \label{eq:genphic}
 \\
 \langle \pi_i \, \pi_j \rangle &=&  \frac{1}{2}\left( K^{1/2} \right)_{ij}  \label{eq:genpic}
\end{eqnarray}

The idea now is to take the two $N\times N$ matrices defined in Eqs (\ref{eq:genphic}),(\ref{eq:genpic}) and truncate them to the
upper left $j_{max} \times j_{max}$ sub-blocks; this is the analog of looking at just the left-side oscillator in our previous example.
Using Eq. (\ref{eq:cdef}), compute the $j_{max} \times j_{max}$  matrix $c$, and then use Eq. (\ref{eq:Swithc}) to compute the
entanglement entropy. Remember that at some point you need to restore the $l,m$ indices; then when computing
the entropy we sum over $m$, which just gives a factor of $(2l+1)$, and sum over $l$ from 0 to some value $l_{max}$.
I have provided you with a notebook that does the whole calculation.

\begin{figure}[tb]
\centering
\includegraphics[width=0.45\textwidth]{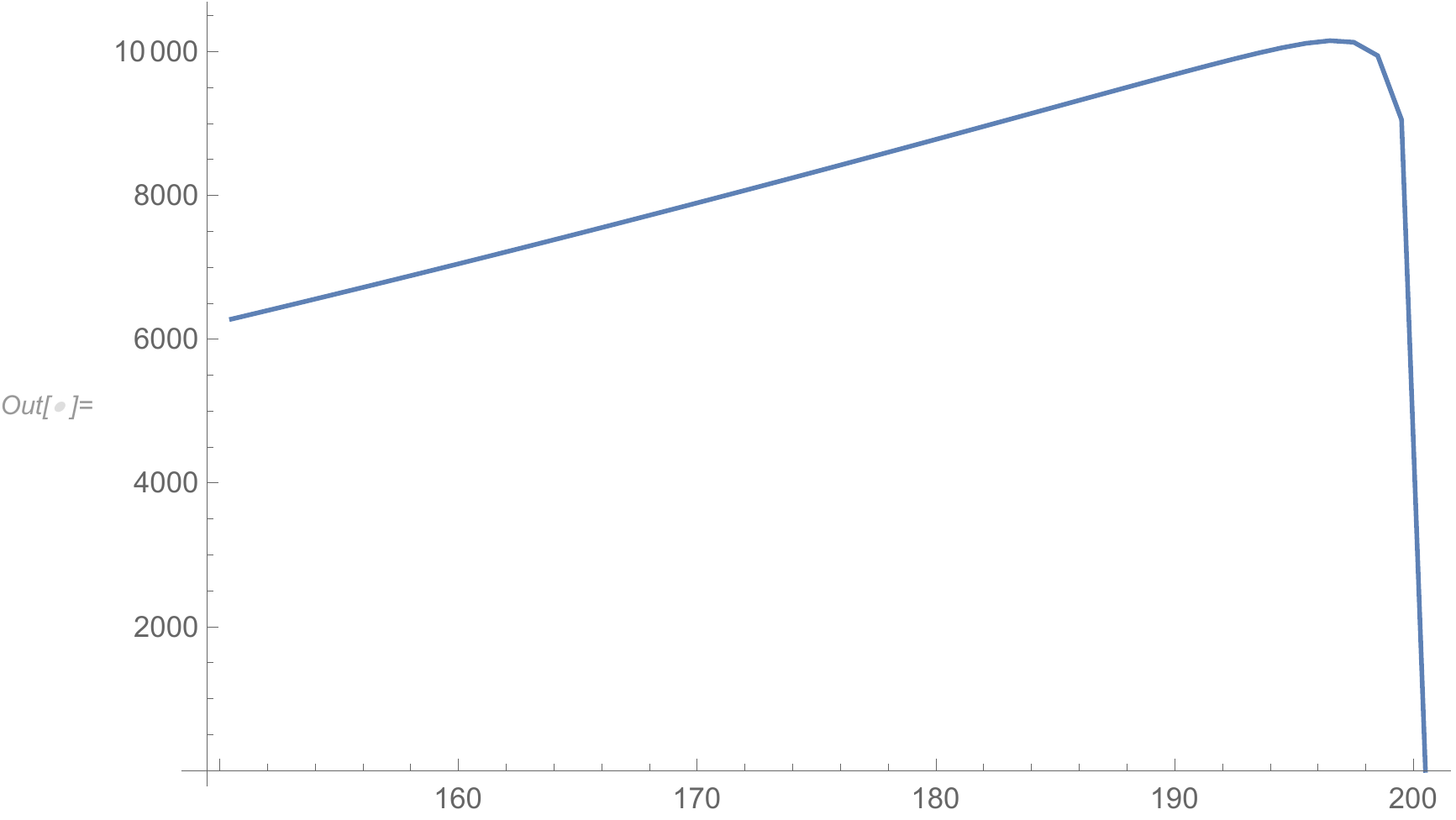}
\includegraphics[width=0.45\textwidth]{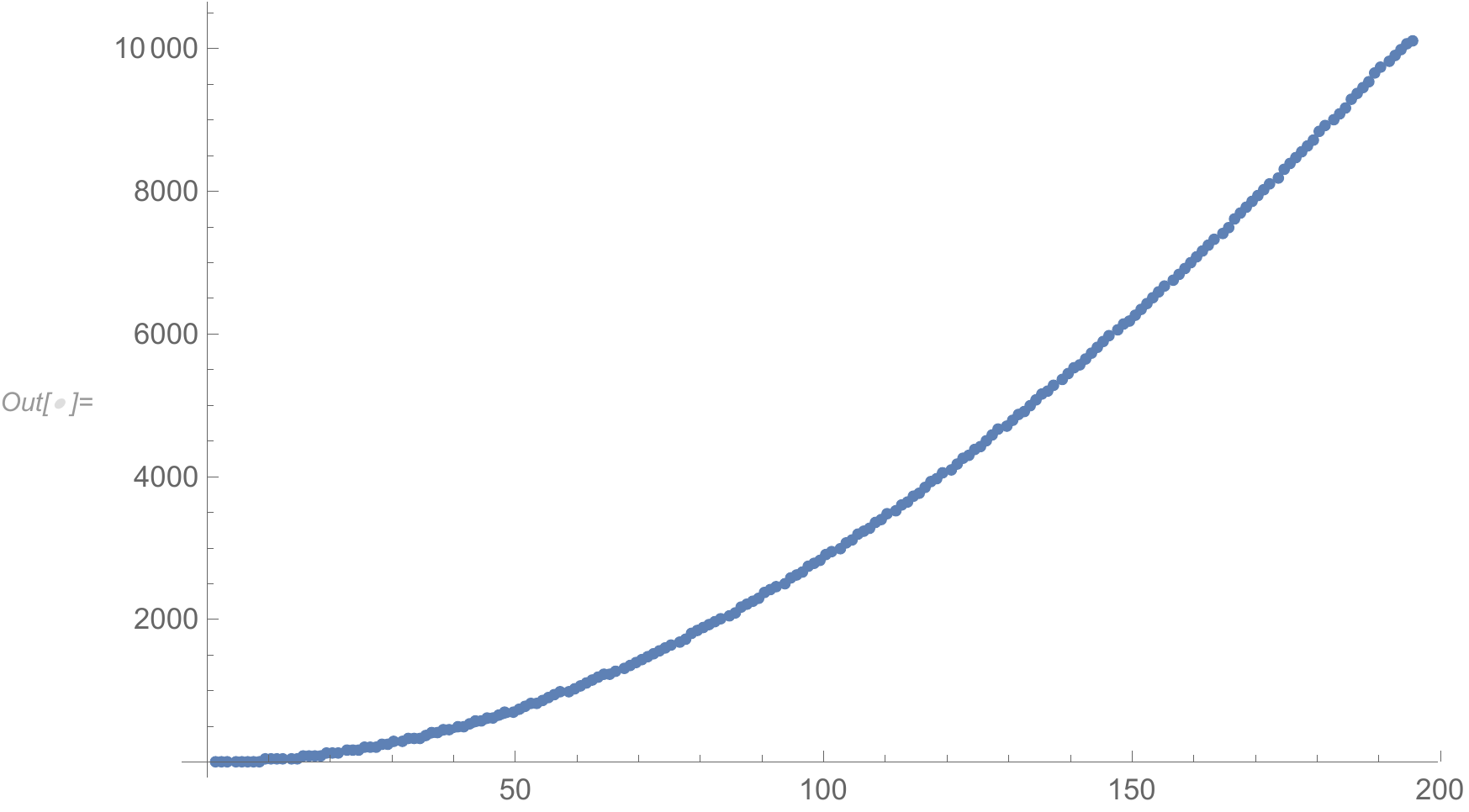}
\caption{Entanglement entropy of oscillators $A$ in a 3d spherical shell, with respect to tracing out oscillators $B$ inside the inner sphere.
Here we have taken the larger radius $R = (200 + 1/2)a$, summed the spherical harmonics up to $l_{\max} = 1000$, and plotted
the entropy $S(\rho_A(r))$ for the inner radius $r$ varying between $0.75 R$ and $R$ (left). In the right figure we show
a fit to the formula $S(\rho_A(r)) = \lambda r^2$ using the computed values of the entropy for $0 < r < 0.975 R$.
The best fit for the constant $\lambda$ is 0.27.   \label{fig:area}}
\end{figure}

Figure (\ref{fig:area}) shows the results for the larger radius $R = (200 + 1/2)a$, summing the spherical harmonics up to $l_{\max} = 1000$, and plotting
the entropy $S(\rho_A(r))$ for the inner radius $r = (j_{max} + 1/2)$ varying between $0.75 R$ and $R$. Notice that entropy vanishes both for $r=0$ (nothing was
traced out) and $r=R$ (everything was traced out). Away from these endpoints the entropy is a smooth monotonic function.
In the right figure we see the result of 
a fit to the formula $S(\rho_A(r)) = \lambda r^2$ using the computed values of the entropy for $0 < r < 0.975 R$.
The best fit for the constant $\lambda$ is 0.27. The agreement of the fit to the computed points is excellent.

We have thus discovered the area law for entanglement entropy for systems of quantum oscillators in $3+1$ dimensions. As you might imagine this property appears
to be rather general, for example it is also true for systems of fermions. There is some literature in exploring the subleading behavior as well. Of course there is
a huge literature in particle physics looking at such phenomena from the point of view of black holes and the famous {\it Beckenstein-Hawking entropy formula}
$S_{\rm BH} = (1/4)A/\ell_{\rm Pl}^2$, where $A$ is the area of the black hole horizon and $\ell_{\rm Pl} = \sqrt{G\hbar/c^3}$ is the Planck length.
However Sredicki's result on the face of it has nothing to with black holes and appears to reflect a general basic property of quantum mechanics
including quantum field theory. More recently a very similar relation called the {\it Ryu-Takayanagi formula} has been much-explored in the context of
AdS/CFT; at least in this context it appears that the black hole properties are related to more general properties of quantum field theories.
What does this all imply for the big picture of how we understand quantum field theory at a fundamental level? That is for you to figure out.

%
%

\section{Simulating quantum field theory on a quantum computer}\label{sec:seven}

One of the big challenges for the next 5-10 years is to identify which problems (or parts of problems) in particle physics most lend themselves to
solution via quantum computing; as you might imagine, those developing quantum processors are also very interested in where a {\it quantum advantage}
might start to manifest itself in real-world scientific problems. This is going to be a long road with many steps and many opportunities; another nice feature
is that there is a lot of overlap with similar challenges in nuclear physics, condensed matter physics, and beyond.

\subsection{Scalar field theory}\label{ss:scalarft}
Having already started to discuss discretizing a scalar field theory, let's ask what we would have to do to actually simulate a scalar field theory on a
quantum computer. In the previous section we already showed how to treat a scalar quantum field theory in terms of a finite number of quantum oscillators. We could just
as well add a $\phi^4$ interaction and generalize Eq. (\ref{eq:HK}) to 
\begin{eqnarray}
H = \frac{1}{2} \sum_{i=1}^{N} \pi_i^2 + \frac{1}{2}  \sum_{i,j=1}^{N} \phi_i K_{ij} \phi_j + \frac{\lambda}{4!} \sum_{i=1}^{N} \phi_i^4
\label{eq:HK2}
\end{eqnarray}
where here it is understood that $K_{ij}$ encodes both the spatial gradient terms and a possible mass term. 

This is still not good enough to map the dynamics onto a quantum computer made of qubits, since each $\phi_i$ can take arbitrary real values
(or complex values if we generalize to a charged scalar); to put it another way, each $\phi_i$ has an infinite number of energy eigenstates,
whereas each qubit can only account for two states. 

To get a better handle on this kind of issue, let's think about why you might go to the trouble of simulating a quantum field theory on a quantum
computer in the first place. If the theory is weakly coupled and can be understood in perturbation theory, or if observables of interest can be
computed by Euclidean lattice methods, then you should certainly use these conventional methods. But, as we already mentioned in the introduction,
a lot of fundamental physics involves real time dynamics of strongly-coupled systems, where a quantum advantage seems possible. Even for such
cases, using a quantum computer only makes sense if you find a way to sensibly map the quantum physics of interest to the quantum physics of the
qubits in your computer; this is obvious for analog quantum simulators but equally true for fully programmable digital quantum computers.
Thus for example, you need to take advantage of the fact that quantum field theory Hamiltonians are local: in the discretized language the self-interactions
involve a single site, and the spatial gradients only involve nearest-neighbor couplings. If your mapping to qubits does not preserve this feature then you
may be doomed from the start. Of course you also need to worry about how to minimize errors from discretization of the problem, especially in the NISQ
era where we have a very limited number of qubits available.
Beyond this, you need to think about how your time evolution operator exp($-iHt$) will operate once translated to qubits;
ideally the time evolution in terms of qubits will traverse the Hilbert space of qubits in a way that is not significantly more complex than what happens
in the Hilbert space of the original system. These kinds of considerations require new ways of thinking about even the most familiar systems in
particle physics \cite{Jordan:2011ne,Jordan:2011ci}.

Going back now to our discretized scalar field theory, suppose we try to map it to a finite number of qubits by truncating the tower of energy eigenstates at each
site. Since we don't know the eigenstates of the strongly-coupled interacting theory, we would have to do something like truncate the tower of eigenstates of the
free Hamiltonian, then turn on the $\lambda$ self-interaction but restricted to the low-lying eigenstates of the free theory. It should not be a surprise that this
introduces unacceptably large discretization errors. A more clever approach is to introduce extra auxiliary parameters into the matrix $K_{ij}$, then try
to adjust those parameters to minimize the discretization errors that appear when you turn on $\lambda$. This approach fails for a different reason,
which is that now the time evolution operator moves you around  in the artificial truncated Hilbert space, which has no simple relation to the original
Hilbert space. You thus find that time evolution immediately entangles all of your qubits to all of your other qubits \cite{Klco:2018zqz}.

The better answer is to {\it digitize} each of the single site fields $\phi_i$, i.e. to only allow them to take a fixed number of values parametrized by integers 
that vary between 0 and $N_\phi$$-$$ 1$, where $N_\phi = 2^{n_q}$ and $n_q$ is the number of qubits that you can afford to represent the field at each site.
Suppressing the site index we can write:
\begin{eqnarray}
\phi = \Delta \left( n_\phi -  \frac{N_\phi}{2} \right) \; ; \quad n_\phi = 0,1,2,\ldots N_\phi - 1
\end{eqnarray}
For example, suppose we only want to use three qubits to denote each $\phi$. Then we require $\phi$ to only take the values such that
\begin{eqnarray}
\phi_q \equiv \left( \frac{2\phi}{\Delta} + 1 \right) \in
[-7,-5,-3,-1,1,3,5,7]
\end{eqnarray}
where I have introduced a rescaled shifted version of $\phi$ denoted $\phi_q$ that is more qubit friendly.
This means that $\phi_q$ as an operator has the following effect on the states of the three qubits:
\begin{eqnarray}
\phi_q \ket{000} &=& 7\ket{000} \nonumber\\
\phi_q \ket{001} &=& 5\ket{001} \nonumber\\
\phi_q \ket{010} &=& 3\ket{010} \nonumber\\
\phi_q \ket{011} &=& 1\ket{011} \\
\phi_q \ket{100} &=& -1\ket{100} \nonumber\\
\phi_q \ket{101} &=& -3\ket{101} \nonumber\\
\phi_q \ket{110} &=& -5\ket{110} \nonumber\\
\phi_q \ket{111} &=& -7\ket{111} \nonumber
 \label{eq:qphi}
\end{eqnarray}
You should check that we can represent the scalar field operator on these qubits in terms of Pauli matrices as
\begin{eqnarray}
\phi_q =
4\sigma_2^z \otimes \mathbbm{1}_1 \otimes \mathbbm{1}_0 +  2\mathbbm{1}_2 \otimes \sigma_1^z \otimes \mathbbm{1}_0 +  \mathbbm{1}_2 \otimes \mathbbm{1}_1 \otimes \sigma_0^z
\end{eqnarray}
We also need a realization of $\phi^2$ and $\phi^4$, since these appear in the Hamiltonian. You can check that the following works for $\phi_q^2$:
\begin{eqnarray}
\phi_q^2 = 
16\, \sigma_2^z \otimes \sigma^z_1 \otimes \mathbbm{1}_0 +  8\, \sigma^z_2 \otimes \mathbbm{1}_1 \otimes \sigma^z_0 + 4\, \mathbbm{1}_2 \otimes \sigma^z_1 \otimes \sigma_0^z
+21\, \mathbbm{1}_2 \otimes  \mathbbm{1}_1 \otimes \mathbbm{1}_0 
\end{eqnarray}

You may object at this point that such an extreme discretization cannot possibly retain a useful amount of information about the 
original quantum system. Let's see how well we are doing for the case of the free theory, where each $\phi(t)$ is just a harmonic oscillator
and $\pi(t)$ is the conjugate: eigenstates of $\phi$ are like position eigenstates with eigenvalues that we can call $x$, and eigenstates of $\pi$ are like momentum eigenstates
with eigenvalues that we can call $p$. In this language we write the energy eigenfunctions and their Fourier transforms as
\begin{eqnarray}
\Psi_n(x) = \frac{1}{\sqrt{2^n n! \sqrt{\pi}}} {\rm e}^{-x^2/2} \, H_n(x) \; ; \quad
\tilde{\Psi}_n(p) = \frac{(-i)^n}{\sqrt{2^n n! \sqrt{\pi}}} {\rm e}^{-p^2/2} \, H_n(p) 
\label{eq:hermite}
\end{eqnarray}
where the $H_n$ are Hermite polynomials.

By taking superpositions of the 8 states defined in (\ref{eq:qphi}), we can only construct the equivalent of sampling the eigenfunctions in
(\ref{eq:hermite}) at 8 points in "space" (this is really field space). This indeed does not seem like much information. But the eigenfunctions
are exponentially peaked around zero (i.e. zero field value), so to an exponentially good approximation I can truncate them on either side
at some value $x=\pm L$ and then just sample them inside this finite interval. Furthermore, since their Fourier transforms are also strongly peaked,
I can truncate them in a similar way, which is the same thing as saying that I represent the eigenfunctions by a Fourier series with a finite number of
terms. Thus to exponential accuracy I can indeed represent the eigenfunctions by sampling them at just enough points to determine the 
coefficients of the truncated Fourier series. The fancy way of saying this is that I am invoking the {\it Nyquist-Shannon sampling theorem}
\cite{Macridin:2018gdw,Macridin:2018oli}.

\begin{figure}[tb]
\centering
\includegraphics[width=0.75\textwidth]{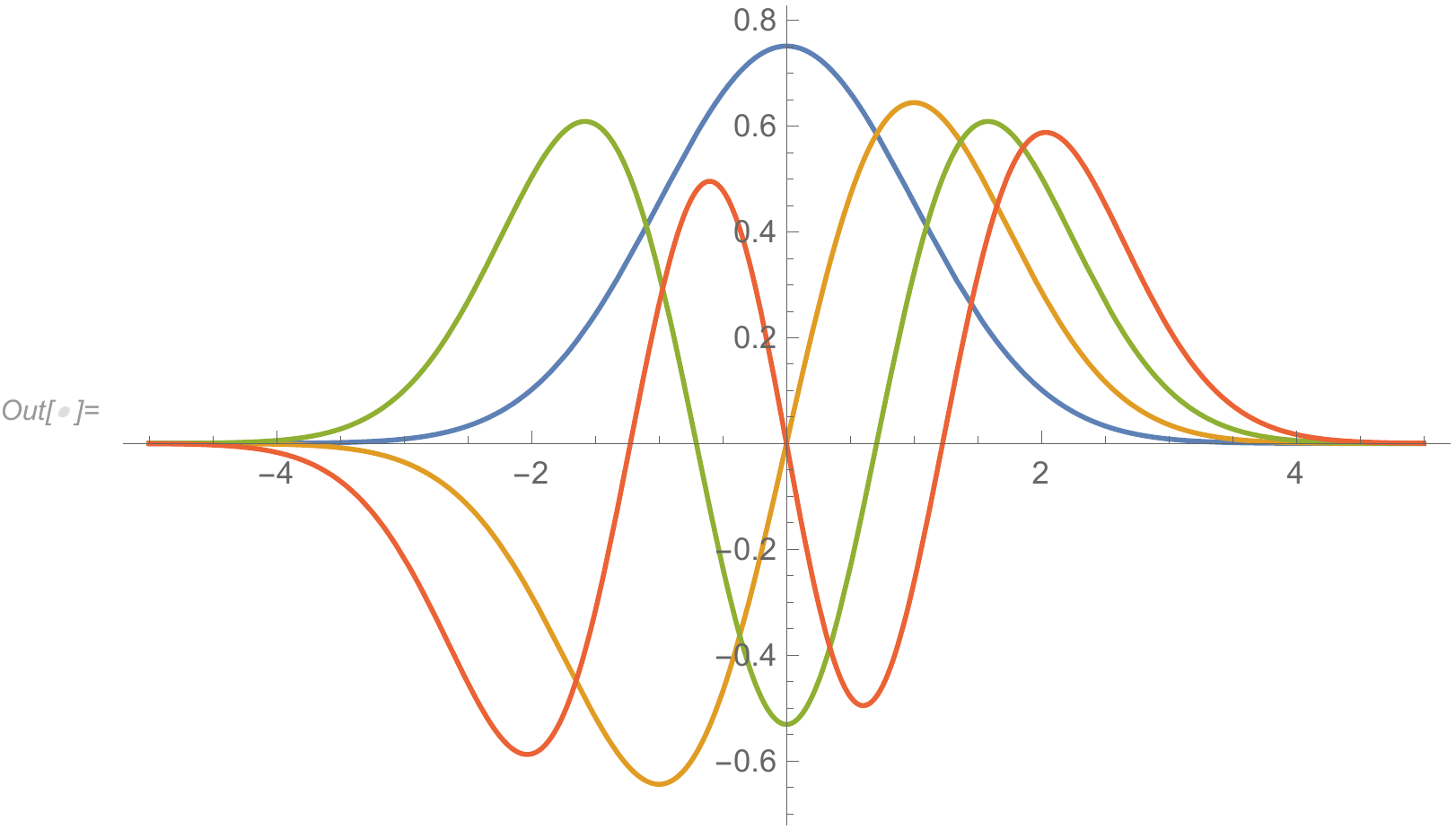}
\caption{The four lowest energy eigenfunctions of a harmonic oscillator.  \label{fig:Hermite}}
\end{figure}

It turns out that in our notation the optimal value for the truncation is $L = \sqrt{N_\phi \pi/2}$. Looking at the four lowest energy eigenfunctions
plotted in Figure (\ref{fig:Hermite}), you see by eye that for $n_q = 3$, $N_\phi = 2^3 = 8$, $L = 3.54$, our 3-qubit representation of the scalar field at each spatial
site really only captures the information in the first couple of eigenfunctions; however things improve very rapidly by adding more qubits. In the notebook you
can look at the case  $n_q = 5$, $N_\phi = 2^5 = 32$, $L = 7.09$, where the sampling captures $>5$-digit precision in the eigenfunctions for the first 16
states. This is not guaranteed to work as well in the interacting theory with $\lambda$ nonzero, but the basic trick is fairly robust.

%
%

\subsection{Gauge theories}\label{ss:Schwinger}
\begin{figure}[tb]
\centering
\includegraphics[width=0.65\textwidth]{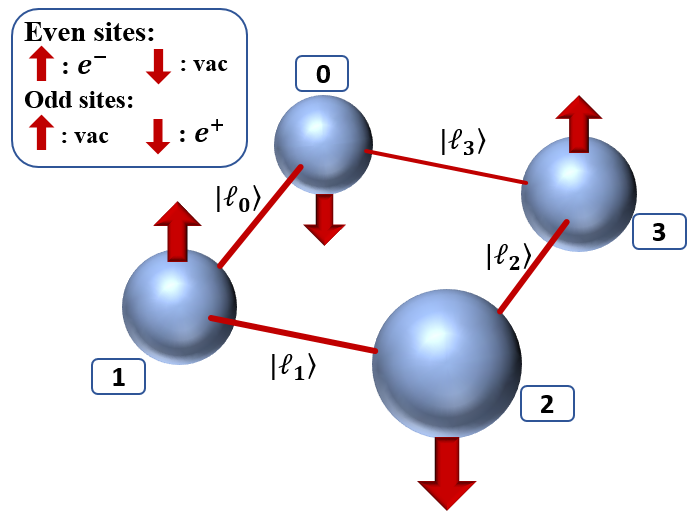}
\caption{Schematic of the qubit and electric flux link structure of the two-site Schwinger model.
Taken from \cite{Klco:2018kyo}.  \label{fig:Schwinger}}
\end{figure}

The simplest example of simulating a gauge theory on a quantum computer starts with the Schwinger model, a representation of
electrodynamics in 1+1 spacetime dimensions that shares some features of 3+1 dimensional QCD, such as confinement and spontaneous breaking
of chiral symmetry. The simplest discretization starts with a periodic spatial lattice with two sites \cite{Klco:2018kyo}, which we double to four sites in order to use a
staggered fermion representation. We label the sites 0,1,2,3, with the understanding that sites 0,2 can be occupied (or not) by an $e^-$, and sites
1,3 can be occupied (or not) by an $e^+$. Using a single qubit vector basis (1 0), (01) to keep track of the fermion occupation of each site, it is convenient to let 
(10) denote the occupied state for sites 0,2, but let (01) denote the occupied state for sites 1,3, as shown in Figure \ref{fig:Schwinger}.
In between sites we may have one or more integer units of electric flux (there is no magnetic flux in a single spatial dimension). Gauss' Law tells us that
the the flux should increase by one unit when moving past a site occupied by an $e^+$, and decrease by one unit when moving past a site occupied by an $e^-$.
To represent the flux with a finite number of qubits we will truncate the integer flux number to only take values 0 and $\pm$1.

%
%

Using the single qubit vector basis to describe the fermion content, the Hamiltonian can be written:
\begin{eqnarray}
H = x \sum_{n=0}^{3} \left( \sigma^+_n \, L^+_n \, \sigma^-_{n+1} + \sigma^+_{n+1} \, L^-_n \, \sigma^-_n \right)
+ \sum_{n=0}^{3} \left( \ell_n^2 + \frac{\mu}{2} (-)^n \sigma^z_n \right)
\label{eq:HSchwinger}
\end{eqnarray}
where $x$ and $\mu$ are dimensionless parameters related to the gauge coupling $g$, electron mass $m$, and lattice spacing $a$ by
$x = 1/(ag)^2$ and $\mu = 2m/(ag^2)$. Here $L^\pm_n$ are link flux lowering and raising operators, and $\ell_n$ is the integer flux number.
Thus the first two terms of the Hamiltonian represent the fermion kinetic energy, the term $\ell_n^2$ is the discretized value of the electric field
energy $E^2$, and the final term is the fermion mass. In general, representing 1+1 dimensional fermion dynamics by Pauli matrices and a qubit basis involves something
called a Jordan-Wigner transformation, to get the correct fermion anticommutation relations, but we will skip the details here \cite{Sachdev}.

\begin{figure}[tb]
\centering
\includegraphics[width=0.65\textwidth]{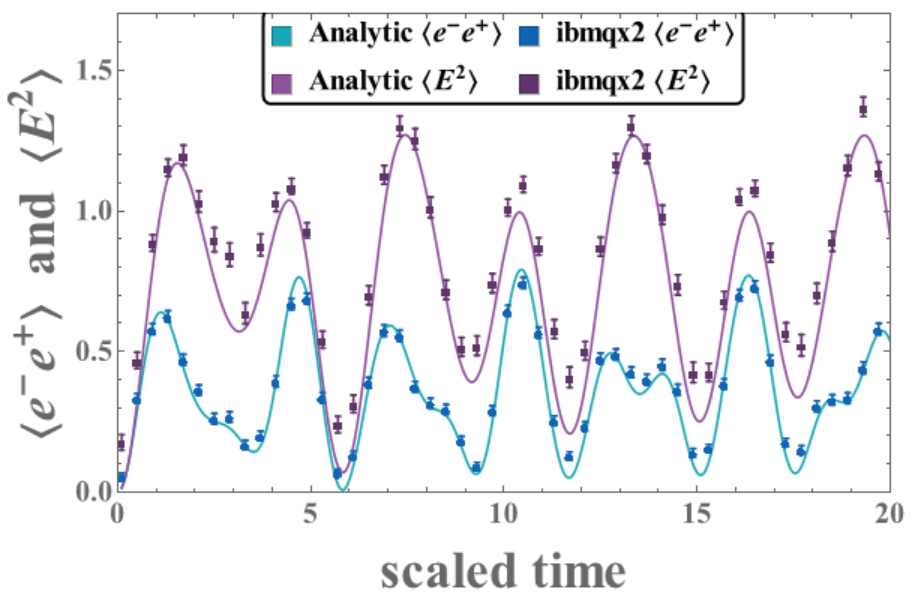}
\caption{Real-time dynamics of the two-site Schwinger model. The initial state has no fermions and no flux (the naive vacuum).
Taken from \cite{Klco:2018kyo}.  \label{fig:SchwingerSim}}
\end{figure}

Imposing Gauss' Law by hand, and truncating the spectrum such that not only does $\ell_n$ only take values 0,$\pm$1, but also $\sum_n \ell^2_n < 4$,
it is not difficult to see that there are a total of four states that have zero charge, zero momentum (and are thus translationally invariant),
and are also even under charge conjugation. We can write these four states as
\begin{eqnarray}
\ket{s_1} &=& \ket{\cdot\cdot\cdot\cdot}\ket{0000} \nonumber \\
\ket{s_2} &=& \frac{1}{2} \left( \ket{e^-e^+\cdot\cdot}\ket{-10000} + \ket{\cdot\cdot e^-e^+}\ket{00-10}
+\ket{e^-\cdot\cdot e^+}\ket{0001} + \ket{\cdot e^+e^-\cdot}\ket{0100} \right) \nonumber\\
\ket{s_3} &=& \frac{1}{\sqrt{2}} \left( \ket{e^-e^+e^-e^+}\ket{-10-10} + \ket{e^-e^+e^-e^+}\ket{0101}  \right) \\
\ket{s_4} &=& \frac{1}{2} \left( \ket{e^-e^+ \cdot\cdot}\ket{0111} + \ket{\cdot\cdot e^-e^+}\ket{1101}
+\ket{e^-\cdot\cdot e^+}{-1-1-10} + \ket{\cdot e^+e^- \cdot}\ket{-10-1-1} \right) \nonumber
\label{eq:Sstates}
\end{eqnarray}
where we have used a double ket notation where the first ket of each pair indicates the fermion occupation and the second ket of each pair
the flux integers. We recognize at this point that what we were initially calling 1+1 dimensional electrodynamics, a $U(1)$ gauge theory, has
in the end been truncated down to the equivalent of a two-site $Z_2$ gauge theory, the simplest possible gauge model.
The state $\ket{s_1}$ is the naive Fock vacuum (no fermions, no flux); consulting the original Hamiltonian Eq. (\ref{eq:HSchwinger}) we see
that it acts on this 4-dimensional truncated subspace as
\begin{eqnarray}
\begin{pmatrix}
-2\mu & 2x & 0 & 0\\
2x & 1 & \sqrt{2}x & 0 \\
0 & \sqrt{2}x & 2 + 2\mu & \sqrt{2}x \\
0 & 0 & \sqrt{2}x & 3
\end{pmatrix}
\label{eq:H4}
\end{eqnarray}
Thus the true ground state of the theory will be a chiral condensate of $e^-e^+$ pairs, i.e. a superposition of the four states in Eq. (\ref{eq:Sstates}).
Since it is not difficult to diagonalize Eq. (\ref{eq:H4}), we don't need a quantum computer to explore this dynamics, but simulating
it on a quantum computer is a good warm up to more challenging gauge theory simulations. Figure \ref{fig:SchwingerSim} shows real-time
dynamical evolution results
obtained by Klco et al. \cite{Klco:2018kyo} running on an IBM quantum computer, versus the analytic results obtained from diagonalizing
Eq. (\ref{eq:H4}). Notice that because of all the truncations and simplifications used here only two qubits are needed for the simulation.
The point of the figure is that even this most simple of gauge theories has nontrivial quantum dynamics, since starting with the naive vacuum
state the unitary time evolution samples the full Hilbert space of states with the same conserved quantum numbers. Thus on a classical computer
this simulation becomes exponentially more difficult as the number of relevant states increases.

%
%

Obviously this simple example is very far away from doing a real-time simulation of QCD. Constructing the pathway to more interesting realistic
models is an area of active research. For example Lamm et al. \cite{Lamm:2019bik}-\cite{Lamm:2019uyc} developed an efficient quantum circuit
to simulate the simplest 2+1 dimensional non-abelian gauge theory, based on the discrete nonabelian group $D_4$, simulated on a two-plaquette lattice.
Their circuit requires 14 qubits and approximately 200 entangling gate operations per time step; circuits with this gate depth are still somewhat beyond
the capabilities of available quantum processors.

\section{Final thoughts}

At the beginning of 2017, Fermilab did not have a quantum science program. Today in 2020 we have SQMS \cite{SQMS} a National QIS Research Center;
we lead a national QIS theory consortium; we have an algorithms group working on quantum computers to explore applications of quantum machine learning,
and we are helping to build and host the world's leading quantum teleportation system \cite{Spiropulu:2020}. We also host  multiple groups developing 
quantum sensor technologies for new kinds of particle physics experiments with unprecedented sensitivities. Two of these collaborations are already
deploying experiments: Dark SRF, a ``light shining through walls" dark photon experiment using ultra-high-Q SRF cavities as photon emitters and receivers,
and MAGIS-100, a 100 meter scale cold atom gradiometer with sensitivity to both dark matter and gravity waves. Particle theorists are playing
important roles in many of these efforts, and I would especially encourage students and postdocs to explore the opportunities in these rapidly
developing areas, which are expanding the traditional boundaries of our field.

As I have tried to indicate in these lectures, just because quantum mechanics has been around for a long time does not mean that we understand it
or are thinking about it in the most productive way. The same goes for particle physics, where even the notion of fundamental ``particles" is starting
to look like a sloppy way of thinking. Consider the electron; it has a mass, but apparently no size, and its mass is apparently nothing more than
an interaction with an invisible Higgs field. The electron carries intrinsic angular momentum, but is not actually spinning; the electron carries charge,
but the effects of that charge are mixed up with properties of the quantum vacuum. The wave function of your electron may be localized away from
the wave function of my electron, but they are both excitations of the same quantum field, and furthermore may be strongly entangled. And of course
the electron is the easy case; most fundamental particles are unstable, and thus ``virtual", or are confined, and thus only indirectly observable. So is it
really more ``physical" to think of the electron as a particle, rather than as a bundle of quantum information?

\section*{Notes on references}
I have borrowed from the excellent general introduction to QIS provided in the book by 
Eleanor Rieffel and Wolfgang Polak {\it ``Quantum Computing: A Gentle Introduction"} \cite{Rieffel}.
Another comprehensive general introduction is the lecture notes by John Preskill \cite{preskill}.
For the Cirq experiments
I have adapted some examples from the \href{https://cirq.readthedocs.io/en/stable/}{Cirq website} and from Jack Hidary's book {\it ``Quantum Computing: An Applied Approach"} \cite{Hidary}.
In Section \ref{sec:four},
I have used the beautiful description of classical measurement given in {\it ``Quantum Measurement and Control"}, the book by Howard M. Wiseman
and Gerard J. Milburn \cite{Wiseman}.
In Sections \ref{ss:decohere} and  \ref{sec:measure},
I have borrowed heavily from the excellent discussion of quantum decoherence given in ``The quantum-to-classical transition and decoherence",  
by Maximilian Schlosshauer \cite{Schlosshauer:2014pgr}.
In Section \ref{sec:six},
I have used the very nice generalization of Srednicki's results presented in lectures by Horacio Casini at the 2016 It from Qubit Summer School \cite{Casini};
for relevant papers see \cite{Calabrese:2004eu}-\cite{Casini:2014yca}.

\section*{Acknowledgments}
These lectures have directly or indirectly benefited from the insights of a number of my colleagues, including Marcela Carena, Dan Carney, Aaron Chou, 
Farah Fahim, Anna Grassellino, Roni Harnik, Hank Lamm, John Preskill, Martin Savage, Panagiotis Spentzouris, and Maria Spiropulu.
Much of the recent research discussed here was supported by the Department of Energy QuantiSED program. 
This manuscript has been authored by Fermi Research Alliance, LLC under Contract No. DE-AC02-07CH11359 with the U.S. Department of Energy, Office of Science, Office of High Energy Physics.



\begin{thebibliography}{99}
 
 \bibitem{notebooks}
 The Jupyter Python notebook and two Mathematica notebooks used in these lectures can be downloaded from Github
 at \href{https://github.com/jlykken/TASI2020-quantum}{https://github.com/jlykken/TASI2020-quantum}. The Jupyter
 notebook is designed to run in the cloud using \href{https://colab.research.google.com/notebooks/intro.ipynb}{Google Colab}.
 
\bibitem{Ellis:1991qj}
R.~K.~Ellis, W.~J.~Stirling and B.~R.~Webber,
{\it QCD and collider physics},
Camb. Monogr. Part. Phys. Nucl. Phys. Cosmol. \textbf{8}, 1-435 (1996)

\bibitem{Page:1993wv}
D.~N.~Page,
``Information in black hole radiation,''
Phys. Rev. Lett. \textbf{71}, 3743-3746 (1993)
doi:10.1103/PhysRevLett.71.3743
[arXiv:hep-th/9306083 [hep-th]].

\bibitem{Gao:2019nyj}
P.~Gao and D.~L.~Jafferis,
``A Traversable Wormhole Teleportation Protocol in the SYK Model,''
[arXiv:1911.07416 [hep-th]].

\bibitem{Harlow:2018fse}
D.~Harlow,
``TASI Lectures on the Emergence of Bulk Physics in AdS/CFT,''
PoS \textbf{TASI2017}, 002 (2018)
doi:10.22323/1.305.0002
[arXiv:1802.01040 [hep-th]].

\bibitem{Harlow:2014yka}
D.~Harlow,
``Jerusalem Lectures on Black Holes and Quantum Information,''
Rev. Mod. Phys. \textbf{88}, 015002 (2016)
doi:10.1103/RevModPhys.88.015002
[arXiv:1409.1231 [hep-th]].

\bibitem{Kockum}
A.F.~Kockum,
``Quantum optics with artificial atoms,"
2014 Ph.D. thesis, 
\href{https://www.researchgate.net/publication/284259345_Quantum_optics_with_artificial_atoms}{available at this link}.

\bibitem{Bennett:1992tv}
C.~H.~Bennett, G.~Brassard, C.~Crepeau, R.~Jozsa, A.~Peres and W.~K.~Wootters,
Phys. Rev. Lett. \textbf{70}, 1895-1899 (1993)
doi:10.1103/PhysRevLett.70.1895


 \bibitem{Spiropulu:2020}
 R. Valivarthi, S. Davis, C. Pe\~{n}a et al,
``Teleportation Systems Towards a Quantum Internet", PRX Quantum 1, 020317 (2020)
doi = {10.1103/PRXQuantum.1.020317},
[arXiv:2007.11157[quant-ph]].

 \bibitem{blueprint:2020}
 P. Dabbar, ``The Quantum Internet of the Future is Here", article and link to the full report
 \href{https://www.energy.gov/articles/quantum-internet-future-here}{on this DOE website}.
 
 \bibitem{Wiseman}
H.M. Wiseman and G.J. Milburn,
{\it Quantum Measurement and Control},
Cambridge University Press (2009).
 
 \bibitem{NC}
M. Nielsen and I. Chuang,
{\it Quantum Computation and Quantum Information},
Cambridge University Press (2000).
 
\bibitem{Zurek:1981xq}
W.~H.~Zurek,
``Pointer Basis of Quantum Apparatus: Into What Mixture Does the Wave Packet Collapse?,''
Phys. Rev. D \textbf{24}, 1516-1525 (1981)
doi:10.1103/PhysRevD.24.1516
 
\bibitem{Zurek:1982ii}
W.~H.~Zurek,
``Environment induced superselection rules,''
Phys. Rev. D \textbf{26}, 1862-1880 (1982)
doi:10.1103/PhysRevD.26.1862
 
\bibitem{Schlosshauer:2014pgr}
M.~Schlosshauer,
``The quantum-to-classical transition and decoherence,''
[arXiv:1404.2635 [quant-ph]].
 
\bibitem{Bell:1964kc}
J.~S.~Bell,
``On the Einstein-Podolsky-Rosen paradox,''
Physics Physique Fizika \textbf{1}, 195-200 (1964)
doi:10.1103/PhysicsPhysiqueFizika.1.195

\bibitem{Clauser:1969ny}
J.~F.~Clauser, M.~A.~Horne, A.~Shimony and R.~A.~Holt,
``Proposed experiment to test local hidden variable theories,''
Phys. Rev. Lett. \textbf{23}, 880-884 (1969)
doi:10.1103/PhysRevLett.23.880
 
\bibitem{Srednicki:1993im}
M.~Srednicki,
``Entropy and area,''
Phys. Rev. Lett. \textbf{71}, 666-669 (1993)
doi:10.1103/PhysRevLett.71.666
[arXiv:hep-th/9303048 [hep-th]].

\bibitem{Maldacena:2001kr}
J.~M.~Maldacena,
``Eternal black holes in anti-de Sitter,''
JHEP \textbf{04}, 021 (2003)
doi:10.1088/1126-6708/2003/04/021
[arXiv:hep-th/0106112 [hep-th]].

\bibitem{Shenker:2013pqa}
S.~H.~Shenker and D.~Stanford,
``Black holes and the butterfly effect,''
JHEP \textbf{03}, 067 (2014)
doi:10.1007/JHEP03(2014)067
[arXiv:1306.0622 [hep-th]].


\bibitem{Jordan:2011ne}
S.~P.~Jordan, K.~S.~M.~Lee and J.~Preskill,
``Quantum Algorithms for Quantum Field Theories,''
Science \textbf{336}, 1130-1133 (2012)
doi:10.1126/science.1217069
[arXiv:1111.3633 [quant-ph]].

\bibitem{Jordan:2011ci}
S.~P.~Jordan, K.~S.~M.~Lee and J.~Preskill,
``Quantum Computation of Scattering in Scalar Quantum Field Theories,''
Quant. Inf. Comput. \textbf{14}, 1014-1080 (2014)
[arXiv:1112.4833 [hep-th]].

\bibitem{Klco:2018zqz}
N.~Klco and M.~J.~Savage,
``Digitization of scalar fields for quantum computing,''
Phys. Rev. A \textbf{99}, no.5, 052335 (2019)
doi:10.1103/PhysRevA.99.052335
[arXiv:1808.10378 [quant-ph]].

\bibitem{Macridin:2018gdw}
A.~Macridin, P.~Spentzouris, J.~Amundson and R.~Harnik,
``Electron-Phonon Systems on a Universal Quantum Computer,''
Phys. Rev. Lett. \textbf{121}, no.11, 110504 (2018)
doi:10.1103/PhysRevLett.121.110504
[arXiv:1802.07347 [quant-ph]].

\bibitem{Macridin:2018oli}
A.~Macridin, P.~Spentzouris, J.~Amundson and R.~Harnik,
``Digital quantum computation of fermion-boson interacting systems,''
Phys. Rev. A \textbf{98}, no.4, 042312 (2018)
doi:10.1103/PhysRevA.98.042312
[arXiv:1805.09928 [quant-ph]].

\bibitem{Klco:2018kyo}
N.~Klco, E.~F.~Dumitrescu, A.~J.~McCaskey, T.~D.~Morris, R.~C.~Pooser, M.~Sanz, E.~Solano, P.~Lougovski and M.~J.~Savage,
``Quantum-classical computation of Schwinger model dynamics using quantum computers,''
Phys. Rev. A \textbf{98}, no.3, 032331 (2018)
doi:10.1103/PhysRevA.98.032331
[arXiv:1803.03326 [quant-ph]].

\bibitem{Sachdev}
S. Sachdev,
{\it Quantum Phase Transitions},
Cambridge University Press (2011).

\bibitem{Lamm:2019bik}
H.~Lamm \textit{et al.} [NuQS],
``General Methods for Digital Quantum Simulation of Gauge Theories,''
Phys. Rev. D \textbf{100}, no.3, 034518 (2019)
doi:10.1103/PhysRevD.100.034518
[arXiv:1903.08807 [hep-lat]].

\bibitem{Alexandru:2019nsa}
A.~Alexandru \textit{et al.} [NuQS],
``Gluon Field Digitization for Quantum Computers,''
Phys. Rev. D \textbf{100}, no.11, 114501 (2019)
doi:10.1103/PhysRevD.100.114501
[arXiv:1906.11213 [hep-lat]].

\bibitem{Lamm:2019uyc}
H.~Lamm \textit{et al.} [NuQS],
``Parton physics on a quantum computer,''
Phys. Rev. Res. \textbf{2}, no.1, 013272 (2020)
doi:10.1103/PhysRevResearch.2.013272
[arXiv:1908.10439 [hep-lat]].

\bibitem{SQMS}
See the website for the Superconducting Quantum Materials and Systems (SQMS) NQI Center at \href{https://sqms.fnal.gov/}{https://sqms.fnal.gov}.


\bibitem{Rieffel}
E~Rieffel, and W~Polak,
{\it Quantum Computing: A Gentle Introduction},
The MIT Press (2011)

\bibitem{preskill}
J. Preskill, 
{\it Quantum Computation},
lecture notes from Caltech course, available on \href{http://theory.caltech.edu/~preskill/ph219/ph219_2018-19}{this website}.

\bibitem{Hidary}
J.~Hidary,
{\it Quantum Computing: An Applied Approach},
Springer (2019).

\bibitem{Casini}
H. Casini,
{\it Entanglement Entropy and QFT},
Lectures at the It from Qubit Summer School, Perimeter Institute, 2016: 
\href{https://www.perimeterinstitute.ca/it-qubit-summer-school/resources/entanglement-qft}{website}.

\bibitem{Calabrese:2004eu}
P.~Calabrese and J.~L.~Cardy,
``Entanglement entropy and quantum field theory,''
J. Stat. Mech. \textbf{0406}, P06002 (2004)
doi:10.1088/1742-5468/2004/06/P06002
[arXiv:hep-th/0405152 [hep-th]].


\bibitem{Casini:2009sr}
H.~Casini and M.~Huerta,
``Entanglement entropy in free quantum field theory,''
J. Phys. A \textbf{42}, 504007 (2009)
doi:10.1088/1751-8113/42/50/504007
[arXiv:0905.2562 [hep-th]].


\bibitem{Casini:2014yca}
H.~Casini, F.~D.~Mazzitelli and E.~Testé,
``Area terms in entanglement entropy,''
Phys. Rev. D \textbf{91}, no.10, 104035 (2015)
doi:10.1103/PhysRevD.91.104035
[arXiv:1412.6522 [hep-th]].



\end{thebibliography}
\end{document}